\documentclass[pdflatex,sn-nature]{sn-jnl}

\usepackage{graphicx}
\usepackage{amsmath,amssymb,amsfonts}
\usepackage{bm}
\usepackage{booktabs}
\usepackage{textcomp}

\usepackage{xcolor}
\usepackage{pdfpages}

\begin{document}

\title[Constant-per-layer-depth matrix product state pretrained ansatz]{Constant-Per-Layer-Depth Matrix Product State Pretrained Ansatz for Noisy Distributed Quantum Processors}

\author[1]{\fnm{Seongpyo} \sur{Hong}}

\author[1,2]{\fnm{Woodo} \sur{Lee}}

\author[1]{\fnm{Yong-Su} \sur{Kim}}

\author*[3,4,5]{\fnm{Seung-Sup B.} \sur{Lee}}\email{sslee@snu.ac.kr}

\author*[1,6]{\fnm{Junghyun} \sur{Lee}}\email{jh\_lee@kist.re.kr}

\affil[1]{\orgdiv{Center for Quantum Technology}, \orgname{Korea Institute of Science and Technology}, \orgaddress{\city{Seoul}, \postcode{02792}, \country{South Korea}}}

\affil[2]{\orgname{xDots}, \orgaddress{\city{Seoul}, \country{South Korea}}}

\affil[3]{\orgdiv{Department of Physics and Astronomy}, \orgname{Seoul National University}, \orgaddress{\city{Seoul}, \postcode{08826}, \country{South Korea}}}

\affil[4]{\orgdiv{Center for Theoretical Physics}, \orgname{Seoul National University}, \orgaddress{\city{Seoul}, \postcode{08826}, \country{South Korea}}}

\affil[5]{\orgdiv{Institute for Data Innovation in Science}, \orgname{Seoul National University}, \orgaddress{\city{Seoul}, \postcode{08826}, \country{South Korea}}}

\affil[6]{\orgdiv{Division of Quantum Information}, \orgname{KIST School, Korea University of Science and Technology}, \orgaddress{\city{Seoul}, \postcode{02792}, \country{Republic of Korea}}}

\abstract{
Distributed quantum processors could scale variational algorithms
beyond single devices, but circuit depth, communication overhead,
and noise limit their performance.
Here we show in simulations that circuit layout affects the
accuracy of a matrix-product-state (MPS) pretrained ansatz under noise, and extend a
construction with constant per-layer depth to modular processors.
We compare ladder, mixed-canonical, and brick-wall realizations with
matched per-layer two-qubit-block resources.
Despite comparable ideal variational quantum eigensolver (VQE) performance
and gradient scales, the constant-depth brick-wall architecture, whose
shallower schedule reduces circuit duration and idle-time decoherence,
yields the lowest noisy energy error and, after zero-noise-extrapolation
(ZNE)-assisted VQE, the lowest noiseless energy error.
On modular processors with one communication qubit per quantum processing
unit (QPU), a nearest-neighbor QPU-path schedule keeps the per-layer depth
constant for three or more QPUs.
In the qubit order fixed by the MPS, some interactions of the target
Heisenberg chain are long-range. Inter-QPU links that realize them improve
connectivity but expose idle qubits to decoherence during communication.
Without error mitigation, circuits with such links are more accurate than
the single-processor brick-wall only when inter-QPU gates are at most twice as slow
and noisy as intra-QPU gates. With ZNE-assisted VQE at short gate durations, they can
remain more accurate even for inter-QPU gates ten times as slow and noisy.
These results support hardware-aware co-design of tensor-network
pretraining, circuit scheduling, and communication topology for
variational quantum computation on noisy modular hardware.
}


\maketitle

\section{Introduction}

Modular distributed quantum processors provide a scalable route to extending
near-term quantum computation beyond the size of a single quantum chip by
interconnecting multiple quantum processing units (QPUs) through
quantum-networking links
\cite{Caleffi2024, Monroe2014_atomic_modular, Main2025_DQC_optical,
Aghaee_Rad2025_modular_photonic, Wehner2018}.
Variational quantum algorithms (VQAs), which optimize parameterized quantum
circuits (PQCs) within a hybrid quantum--classical loop, are natural
candidates for such hardware because their circuit architectures can be
adapted to the available connectivity and depth constraints
\cite{Cerezo2021VQAReview, Khait2023_VQEDQC, Cattelan2025}.
The hardware-efficient variational quantum eigensolver (VQE), for example,
adapts the circuit to the native interactions of the device \cite{Kandala2017HardwareEfficientVQE}, and a
two-qubit VQE has been demonstrated across networked photonic processors
using pre-shared entanglement and local weak measurements
\cite{Lee2026DistributedPhotonicVQE}.
Executing variational circuits, however, incurs two costs that recur at every evaluation of
the ansatz.
First, on noisy intermediate-scale quantum (NISQ) hardware, gate errors and
decoherence accumulate with circuit depth and duration, so that even ansatzes
with strong ideal-circuit performance can become unreliable once executed
\cite{Preskill2018NISQ, Uvarov2024VQE-ZNE, Dalton2024noisyVQE,
Wang2021NoiseBP-VQA, Fontana2021EvalNoiseVQA,
Schumann2024NoiseBPArbitrarylayered}; error-mitigation techniques such as
zero-noise extrapolation (ZNE)
\cite{Temme2017ZNE, LiBenjamin2017EfficientDistributed, Endo2018QEM} can
partially suppress these errors, but finite-shot implementations incur a
statistical overhead that can grow strongly with circuit depth and noise strength
\cite{Cai2023QEMReview, Takagi2022QEMLimits}.
Second, inter-QPU entangling operations require quantum communication,
heralding, and synchronization, and are therefore slower, less faithful, or
both, compared with local gates
\cite{Caleffi2024, Khait2023_VQEDQC, Ang2024ARQUIN}.
Circuit depth and the number of inter-QPU communication rounds are thus
two key execution resources that a hardware-executable variational ansatz
must conserve.

Tensor-network (TN) states provide compact representations of physically
relevant quantum states by explicitly encoding their correlation and
entanglement structure \cite{Schollwock2011_DMRGMPS, Verstraete2008}.
Matrix product states (MPSs) in particular can be mapped onto structured
quantum circuits through sequential state-preparation or
matrix-product-disentangler constructions \cite{Ran2020, Rudolph2024MPSDecomposition,
Schon2005ExactEncoding, Szoldra2026ScalableMPSPreparation}.
Such mappings have served as warm starts for PQCs
\cite{Rudolph2023SynergisticPretraining, Dborin2022MPSPretraining}, including
ZNE-assisted VQE \cite{Sun2025PretrainingZNEVQE}; more recently,
density-matrix-renormalization-group (DMRG) \cite{White1992DMRG} solutions
have been compiled into shallow state-preparation circuits executed on quantum
hardware \cite{Jaderberg2026NormalMPSPreparation, Pennington2026SPT100Qubit}.
Renormalization-group-based MPS preparation on superconducting hardware
further shows how reduced circuit depth can improve robustness to noise
\cite{Scheer2025RG80MPS}.
Parallel-sequential circuits, which interpolate between sequential and
brick-wall layouts, show how favorable preparation layouts depend on the
balance between approximation error, gate errors, and idle errors
\cite{WeiMalz2026ParallelSequential}.
In these state-preparation works the circuits are executed with fixed,
classically optimized parameters, whereas in VQE the parameters are further optimized with device
measurements, which can change this balance.

\begin{figure}[t]
\centering
\includegraphics[width=0.85\textwidth]{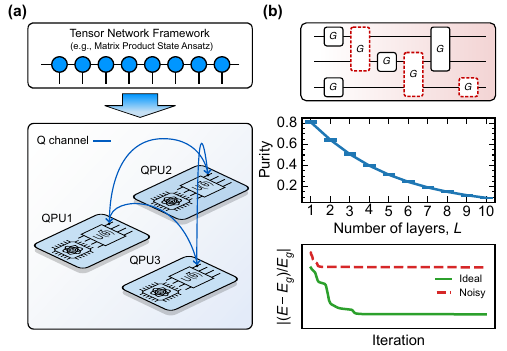}
\caption{Overview of tensor-network--assisted circuit design and the impact of depth-dependent decoherence. 
(a) Schematic workflow: a classically optimized tensor-network state (illustrated here for an MPS) 
is mapped onto a gate-based parameterized circuit and deployed on a 
distributed quantum processor comprising multiple QPUs. 
(b) Illustration of a noisy quantum circuit, 
and representative trends showing purity decay with an increasing number of layers 
and the resulting degradation of variational optimization. 
The purity plot reports the mean over $10{,}000$ random-circuit realizations, 
with error bars indicating one standard deviation, 
whereas the ideal and noisy optimization curves show representative single
optimization trajectories.}
\label{fig1}
\end{figure}

For MPS-pretrained VQE, the choice of layout is consequential: a given target MPS can be
encoded into ladder, mixed-canonical, or brick-wall layouts that use the same
number of two-qubit blocks per layer, yet whose per-layer depths are
$N-1$, $\lceil N/2\rceil$, and $2$ for $N$ qubits.
Circuit realization, rather than state representation alone, is therefore a
central design problem, and MPS pretraining provides a controlled setting in
which one structured target state is mapped to circuits with markedly
different parallel depths and connectivity [Fig.~\ref{fig1}(a)].
Because gradient variance is also governed by local depth
\cite{Zhang2024AbsenceBarrenPlateaus}, a fair comparison of realizations must
additionally control for their ideal-circuit gradient behavior.

When such a circuit is distributed across several QPUs, the same scheduling
problem acquires a communication dimension: two-qubit blocks spanning
different QPUs must be mediated by communication qubits, making the per-layer
depth dependent on the required number of inter-QPU communication rounds.
A circuit architecture with system-size-independent per-layer depth on a
single processor therefore provides a natural basis for constructing
distributed ansatzes that retain shallow local structure while minimizing
communication overhead.
The question for MPS-pretrained circuits is then whether this favorable
scaling persists after QPU partitioning, and how it competes with the
increased variational expressivity enabled by richer inter-QPU
connectivity.

In this work, we study how the circuit realization of MPS pretraining affects
noisy VQE on single and distributed processors.
We compare ladder
\cite{Dborin2022MPSPretraining,Rudolph2023SynergisticPretraining},
mixed-canonical \cite{Dborin2022MPSPretraining}, and brick-wall
\cite{Schuhmacher2025HybridTTN} MPS-pretrained circuits with matched
per-layer two-qubit-block resources. They show comparable ideal convergence
and gradient scales, but under a finite-duration gate-local noise model
the constant-depth brick-wall layout has the lowest noisy error and, after
ZNE-assisted training, the lowest noiseless error. With one communication
qubit per QPU, a nearest-neighbor (NN) QPU-path schedule preserves constant
per-layer depth, whereas all-to-all communication adds blocks and
parameters and its depth grows with the number of QPUs. In two- and
three-QPU simulations of a twisted Heisenberg chain with the qubit order fixed
by the MPS, a two-QPU circuit with a single inter-QPU link (the link circuit) and the three-QPU all-to-all circuit have
lower mean noiseless errors than the brick-wall circuit before and after noiseless VQE, but at the baseline gate duration VQE under noise narrows this advantage.
Without error mitigation they remain more accurate than the brick-wall only
for inter-QPU gates at most twice as slow and noisy as intra-QPU gates. With
ZNE-assisted training at short gate durations, they can remain more accurate
even for gates ten times as slow and noisy. For the distributed circuits,
MPS pretraining mainly accelerates the initial decrease in energy error.
Reordering the twisted model as a nearest-neighbor chain shows that distribution
without a connectivity gain mainly adds communication overhead.

\section{Results}\label{sec:main_results}

\subsection{Structured ansatzes: ideal-circuit and encoding diagnostics}\label{sec:ansatze}

\begin{figure}[t]
\centering
\includegraphics[width=0.85\textwidth]{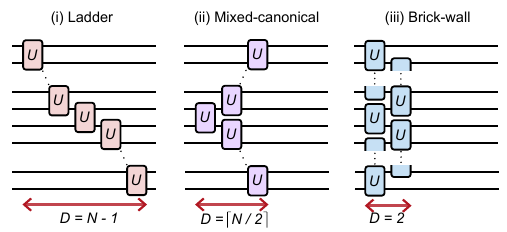}
\caption{Illustration of the three structured ansatzes considered in this work: ladder, mixed-canonical, and brick-wall.
Each layer comprises $N-1$ two-qubit unitary blocks, but the parallelization strategy differs,
resulting in distinct per-layer circuit depths: $D=N-1$ (ladder), $D=\lceil N/2\rceil$ (mixed-canonical), and $D=2$ (brick-wall),
where each brick-wall layer consists of one odd and one even sublayer.
Here $D$ denotes the two-qubit-block depth, written as $D_2$ in Fig.~\ref{fig7}, Methods, and the Supplementary Information.}
\label{fig2}
\end{figure}

We benchmark the three structured circuit families---\emph{ladder}, \emph{mixed-canonical}, and \emph{brick-wall}---under
ideal (noiseless) conditions within a single QPU to compare their VQE energy convergence 
and local gradient-scale behavior on an equal footing.
The three layouts use the same $N-1$ two-qubit unitary blocks per layer but
schedule them with per-layer depths of $N-1$, $\lceil N/2\rceil$, and $2$,
respectively (Fig.~\ref{fig2} and Table~\ref{tab:ansatz_summary}; see Methods).
Unless stated otherwise, the simulations target the Pauli-normalized twisted
Heisenberg chain of $N$ qubits [Eq.~\eqref{eq:twisted_hamiltonian}, Methods],
the open-boundary Heisenberg chain in a permuted qubit order, whose exact
ground energy $E_g$ is obtained by exact diagonalization.
A circuit of $L$ layers has its first $K$ layers initialized by encoding a
bond-dimension-$\chi$ MPS approximation of the target ground state, with all
parameters trainable during VQE, and we report the relative energy error
$|(E-E_g)/E_g|$.

\begin{table}[t]
\caption{
Comparison of the structured ansatzes considered in this work.
For each ansatz, we report the per-layer circuit depth and the encoding state fidelity
$F=|\langle\psi_{\mathrm{MPS}}|\psi_{\mathrm{circuit}}\rangle|^2$
for bond dimensions $\chi=2$ and $\chi=8$.
The target MPSs are bond-dimension-constrained
DMRG approximations to the ground state of the Pauli-normalized twisted
Heisenberg Hamiltonian studied throughout this work, here for $N=8$.
For $\chi=2$, the encoding fidelity is reported after encoding with a single circuit layer ($K=1$).
For $\chi=8$, the encoding fidelities are reported after encoding with $K=1$, $2$, and $3$ layers.
Circuit depth is measured in units of two-qubit unitary blocks, assuming disjoint
blocks are executed in parallel.
For the brick-wall ansatz, one circuit layer comprises one odd and one even sublayer.
}
\label{tab:ansatz_summary}
\renewcommand{\arraystretch}{1.3}
\begin{tabular}{lccc}
\toprule
 & Ladder & Mixed-canonical & Brick-wall \\
\midrule
Depth per layer 
& $N-1$ 
& $\lceil N/2 \rceil$ 
& $2$ \\
\midrule
$\chi=2$ state fidelity ($K=1$) 
& 1.0 (exact) 
& 1.0 (exact) 
& 0.9866 \\
\midrule
$\chi=8$ state fidelity ($K=1$) 
& 0.5635 
& 0.5635 
& 0.4903 \\
$\chi=8$ state fidelity ($K=2$) 
& 0.8136 
& 0.8147 
& 0.8154 \\
$\chi=8$ state fidelity ($K=3$) 
& 0.9904 
& 0.9936 
& 0.9911 \\
\botrule
\end{tabular}
\end{table}

\begin{figure}[t]
\centering
\includegraphics[width=\textwidth]{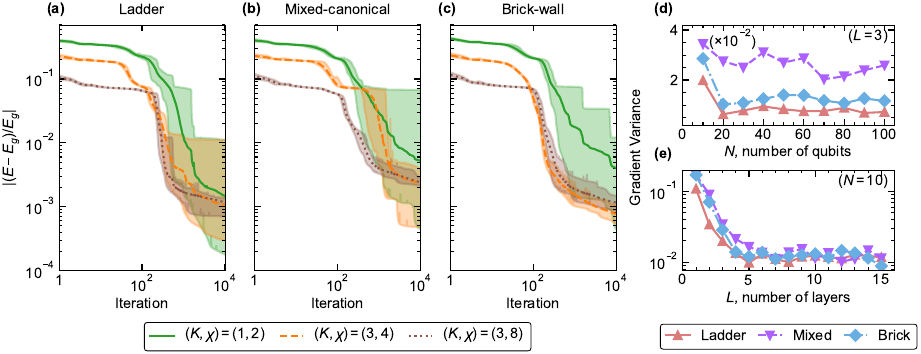}
\caption{Ideal-circuit benchmarks of energy convergence and gradient-variance diagnostics of the three structured ansatzes.
Left: Noiseless VQE convergence of the relative energy error $|(E-E_g)/E_g|$
for the (a) ladder, (b) mixed-canonical, and (c) brick-wall ansatzes
with a fixed total layer number $L=4$. 
Results are shown for different combinations of the number of 
pretrained layers $K$ and the MPS bond dimension $\chi$ (see legend). 
Solid lines denote the mean over $20$ independent
random initializations, and shaded regions indicate the corresponding min--max range (see Methods for initialization details).
Right: Gradient-variance diagnostics, computed as the unbiased sample variance over $100$ independent random parameter realizations. 
(d) shows the gradient variance as a function of system size $N$ with fixed $L=3$, 
whereas (e) shows the gradient variance as a function of the number of layers $L$ for a fixed system size of $N=10$. 
The gradient is evaluated with respect to the \texttt{RYY} (IsingYY) parameter 
in the first two-qubit unitary block according to each ansatz's construction order; 
for the mixed-canonical ansatz, this corresponds to the central two-qubit block.  
The selected parameter is matched by construction order rather than by  
physical bond across the different ansatzes.
}
\label{fig3}
\end{figure}

Fig.~\ref{fig3}(a)--(c) shows the relative energy error for $N=8$, a fixed
total layer number of $L=4$, and several combinations of pretrained layers $K$ and MPS bond dimension $\chi$.
Across the explored $(K,\chi)$ configurations, all three circuit families 
achieve comparable final energy errors and convergence behavior under noiseless conditions.
Increasing $\chi$, $K$, or both generally improves convergence relative to weak pretraining [e.g., $(K,\chi)=(1,2)$], although the dependence is not monotonic for every ansatz and setting.

To complement the energy-based benchmarks, we examine local gradient scales through gradient-variance diagnostics~\cite{McClean2018BarrenPlateaus,Cerezo2021CostFunctionBarrenPlateaus}, which have also been studied for TN-inspired circuit architectures~\cite{CerveroMartin2023BarrenPlateausQTN}.
Fig.~\ref{fig3}(d),(e) reports the gradient variance over $100$ random parameter realizations as a function of $N$ at $L=3$ and of $L$ at $N=10$.
Across the explored depth regime, all three ansatzes exhibit similar gradient-variance trends.  We therefore read Fig.~\ref{fig3}(d),(e) as evidence for comparable gradient scales in the shallow-circuit regime, not as a characterization of the optimization landscape.

\begin{figure}[t]
\centering
\includegraphics[width=0.85\textwidth]{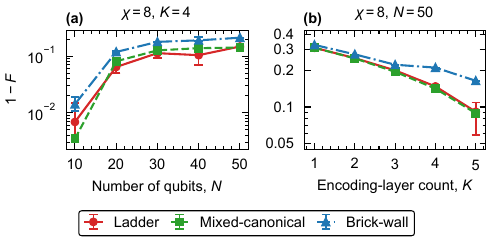}
\caption{Ideal MPS-to-circuit encoding infidelity for the twisted Heisenberg target MPS.
(a) State infidelity, $1-F$, as a function of system size for fixed $\chi=8$ and $K=4$.
(b) State infidelity as a function of the number of encoding layers $K$ for fixed $\chi=8$ and $N=50$.
The fidelity is computed between the target MPS and the corresponding circuit-prepared state for the three structured circuit layouts.
Markers show the mean over six restarts of the encoding
optimization and error bars the min--max range (Supplementary
Information).
Here, $K$ denotes the number of circuit layers optimized to encode the
target MPS; such layers serve as the pretrained layers in the VQE
calculations.
}
\label{fig4}
\end{figure}

The encoding fidelities of Table~\ref{tab:ansatz_summary} and Fig.~\ref{fig4}
complete the ideal-circuit diagnostics. With one encoding layer ($K=1$), the
ladder and mixed-canonical layouts admit exact $\chi=2$ encodings, whereas the brick-wall has no such general
guarantee. For the $N=8$, $\chi=8$ targets, the three fidelities are
comparable at $K=2$ and $3$ (Table~\ref{tab:ansatz_summary}).
Fig.~\ref{fig4} extends the encoding benchmark of Table~\ref{tab:ansatz_summary} to larger $N$ through the state infidelity $1-F$ between the $\chi=8$ target MPS and the circuit-prepared state. 
With the encoding layer count fixed at $K=4$ [Fig.~\ref{fig4}(a)],
the mean state infidelity grows by more than an order of magnitude from $N=10$ to $N=50$ in each layout, reflecting the increasing encoding difficulty at fixed $K$. The brick-wall has the largest mean infidelity throughout. The ladder has the widest optimization spread for $N\ge20$: its ranges at $N=30$ and $40$ overlap, while its $N=50$ range lies above both.
At $N=50$ [Fig.~\ref{fig4}(b)], increasing $K$ from $1$ to $5$
reduces the mean infidelity monotonically for all three layouts, to $0.09$ (ladder and mixed-canonical) and $0.16$ (brick-wall) at $K=5$. The ladder and mixed-canonical layouts improve at nearly the same rate, whereas the brick-wall layout improves more slowly and retains the largest residual infidelity.

\subsection{Noisy VQE and ZNE}\label{sec:noisy_results}

\begin{figure}[t]
\centering
\includegraphics[width=0.85\textwidth]{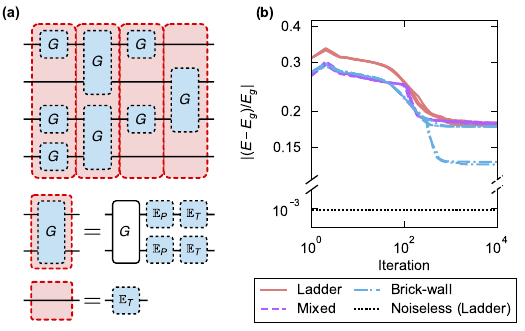}
\caption{Gate noise model and its impact on variational optimization.
(a) Schematic of the noise model used in this work: after each decomposed gate $G$, 
the biased Pauli channel $\mathbb{E}_{P}$ and the thermal-relaxation channel $\mathbb{E}_{T}$
are applied, while idle intervals are modeled by $\mathbb{E}_{T}$ alone.
(b) Noisy VQE convergence of the relative energy error, $|(E-E_g)/E_g|$,
for the three structured ansatzes at $(L,K,\chi)=(4,3,8)$, five runs each.
The dotted horizontal line, below the axis break, indicates the mean final
relative energy error over 20 noiseless VQE runs of the ladder ansatz at the
same setting, with each noiseless run optimized for 20,000 Adam iterations.
}
\label{fig5}
\end{figure}

Apart from the encoding-fidelity differences, which are small for the $N=8$
targets at $K\ge2$ (Table~\ref{tab:ansatz_summary}) but persist at larger $N$
(Fig.~\ref{fig4}), the three layouts show comparable ideal-circuit energy
convergence and gradient scales; their practical differences emerge
primarily under depth-dependent noise. We benchmark them under the combined gate-local noise model, using global unitary folding and quadratic Richardson extrapolation for ZNE (Methods). Biased Pauli ($\mathbb{E}_P$) and thermal-relaxation ($\mathbb{E}_T$) channels act after each decomposed gate; only $\mathbb{E}_T$ acts during idle intervals [Fig.~\ref{fig5}(a)]. Noise therefore accumulates with circuit duration and depth, reducing purity and degrading optimization [Fig.~\ref{fig1}(b)]. 

\begin{figure}[t]
\centering
\includegraphics[width=0.85\textwidth]{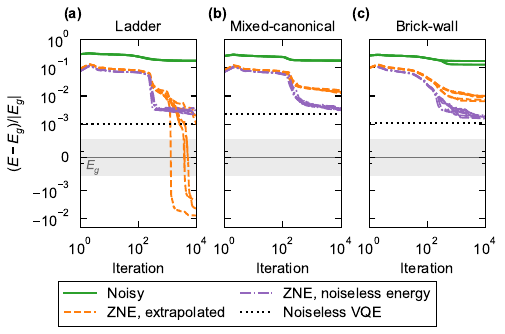}
\caption{
Noisy VQE and ZNE-assisted VQE for the three structured circuit families under the gate-local noise model (Methods) at $(L,K,\chi)=(4,3,8)$. (a) Ladder, (b) mixed-canonical, and (c) brick-wall ansatzes. Each panel shows the signed relative energy error, $(E-E_g)/|E_g|$, as a function of optimization iteration on a symmetric logarithmic axis that is linear in the shaded band between $\pm3\times10^{-4}$; the gray line marks $E_g$. For each ansatz, five independent optimization runs are shown for noisy VQE
(green solid) and for ZNE-assisted VQE, with both the extrapolated energy (orange dashed) and the noiseless energy of the same parameters (purple dash-dotted).
The black dotted line indicates the mean final relative energy error over 20 noiseless VQE runs of the corresponding architecture
at the same setting, with each noiseless run optimized
for 20,000 Adam iterations.
}
\label{fig6}
\end{figure}

Fig.~\ref{fig5}(b) presents five independent noisy-VQE runs for each ansatz at
$N=8$ and $(L,K,\chi)=(4,3,8)$, where the three layouts reach comparable
noiseless VQE errors [Fig.~\ref{fig3}(a)--(c)]; under noise the optimization stalls far above the noiseless VQE error.  At $(L,K,\chi)=(3,2,8)$, by contrast, the
brick-wall reaches a lower noiseless error than the other two layouts, so
the comparison at that setting does not isolate the effect of noise (Supplementary
Information). 
Fig.~\ref{fig6} compares the five noisy-VQE runs of Fig.~\ref{fig5}(b) with five independent ZNE-assisted VQE runs for each circuit family. Gate errors and decoherence often cause an early plateau (green), limiting improvement from further parameter updates. ZNE-assisted VQE reduces the mean final absolute relative error (orange) by more than an order of magnitude. The purple curves show the noiseless energies at the ZNE-trained parameters.

The brick-wall ansatz reaches the lowest mean unmitigated error, $0.159(25)$
versus $0.1802(19)$ for the ladder and $0.1819(5)$ for the mixed-canonical
layout, and all five of its noisy runs end below every ladder and
mixed-canonical run. Its ZNE-trained circuits have the
lowest noiseless errors, $0.0018(2)$ versus $0.0027(6)$ and $0.0034(3)$.
This behavior is consistent with its depth-efficient architecture: 
at equal gate count, the reduced depth shortens the circuit duration and
hence the idle-time decoherence. At the noisy-trained parameters the
noise-induced shift, the difference between the noisy and the noiseless energy
at the same parameters (Sec.~\ref{sec:training_objective}), is $0.108$ for the
brick-wall versus $0.175$ and $0.174$ for the ladder and mixed-canonical
ansatzes, in units of $|E_g|$.
The executed circuits overlap operations from successive layers: at
$(L,K,\chi)=(4,3,8)$ their two-qubit-block depths are $13$, $10$, and $8$
for the ladder, mixed-canonical, and brick-wall layouts, and the ladder
depth exceeds the brick-wall depth by $N-3$ at any $L$, a difference that grows
with the system size (Supplementary Information).
For the ladder ansatz, the extrapolated energy ends below the corresponding
noiseless energy in all five runs and below $E_g$ in three
[Fig.~\ref{fig6}(a)]. Between iterations $1{,}000$ and $10{,}000$ the extrapolated energy
continues to decrease while the noiseless error changes little, so the
optimization lowers the extrapolated energy partly through the
extrapolation error (Sec.~\ref{sec:training_objective}).

The brick-wall retains the lowest mean unmitigated error, the lowest mean
noiseless error after ZNE-assisted training, and the smallest noise-induced
shift when, instead of the separate pretrainings of Fig.~\ref{fig6}
(Methods), noiseless, noisy, and ZNE-assisted VQE start from the same five
initial parameter sets within each layout.
Two further controls at $(L,K,\chi)=(3,2,8)$ are reported in the
Supplementary Information. When the gate error is made dominant over the
relaxation, with a biased-Pauli probability of $5\times10^{-3}$ per active qubit at
$t_{\mathrm{2q}}=5\,\mu$s, the noise-induced shifts of the three layouts no
longer favor the brick-wall, so the smaller shift at the baseline duration
is an effect of the schedule rather than of the gate-local errors.
The brick-wall also has the lowest mean final unmitigated error at every
tested two-qubit-gate duration, $t_{\mathrm{2q}}=5$--$500\,\mu$s.

\subsection{Distributed-QPU ansatz: communication-aware constant-per-layer-depth scheduling}\label{sec:distributed_qpu}

\begin{figure}[p]
\centering
\includegraphics[width=0.60\textwidth]{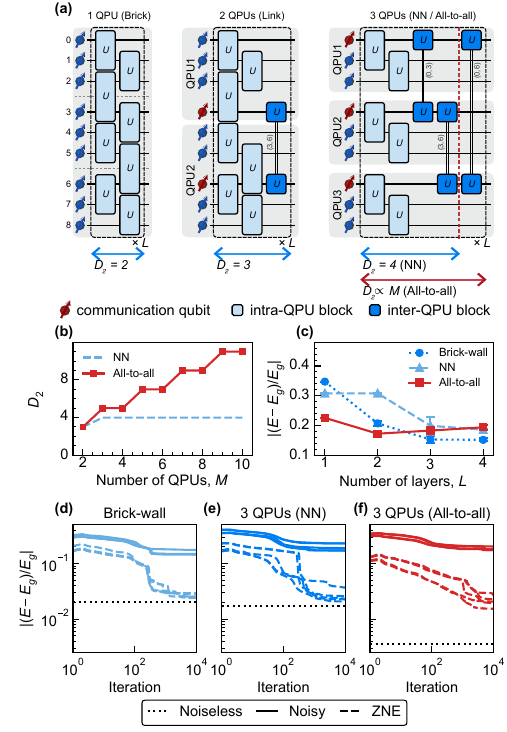}
\caption{
Distributed-QPU extension of the MPS-pretrained ansatz.
(a) One layer of the single-QPU brick-wall circuit on the nine-qubit chain
(left), of the two-QPU link circuit of Fig.~\ref{fig8} (middle), and of the
three-QPU circuits (right). Light-blue blocks form the local brick-wall
pattern within each QPU, dark-blue blocks are the inter-QPU blocks on the
designated communication qubits, and gray dashed lines mark the three-QPU
partition. Inter-QPU links: $(3,6)$ for the two-QPU circuit; $(0,3)$ and
$(3,6)$ for NN connectivity [(e)], and additionally $(0,6)$ for all-to-all
connectivity [(f), right of the red dashed line].
(b) Per-layer two-qubit-block depth $D_2=2+R$ versus the number of QPUs,
$M$, for one communication qubit per QPU, where $R$ is the number of
inter-QPU communication rounds per layer (Supplementary Information).
(c) Mean final noisy-VQE relative energy error over five runs versus the
total layer number $L$ for the brick-wall and three-QPU NN and all-to-all
circuits; error bars give the sample standard deviation [$(K,\chi)$ for
each $L$: Supplementary Information].
(d)--(f) VQE optimization trajectories for $(L,K,\chi)=(3,2,8)$: five
noisy-VQE and five ZNE-assisted VQE runs for the (d) brick-wall, (e)
distributed NN, and (f) distributed all-to-all ansatzes; black dotted
lines mark the mean final noiseless VQE error over five runs of the same
circuit.
(c)--(f) use $t_{\mathrm{2q}}=50\,\mu$s and $(r_t,r_p)=(1,1)$; each run
starts from an independent MPS pretraining, shared by the three methods
in (d)--(f) (Methods).
}
\label{fig7}
\end{figure}

Motivated by the single-processor results, we extend the constant-depth
brick-wall construction to noisy modular processors comprising $M$ interconnected QPUs;
the simulations use a nine-qubit chain ($N=9$), split for the three-QPU
circuits into three QPUs of three qubits with communication qubits $0$, $3$,
and $6$.
Each distributed layer combines local brick-wall blocks within every QPU with
inter-QPU blocks acting on one designated communication qubit per QPU, under
either nearest-neighbor (NN) or all-to-all inter-QPU connectivity
[Fig.~\ref{fig7}(a); Methods]. 
Inter-QPU entangling operations are typically slower and less reliable than local gates \cite{Caleffi2024, Khait2023_VQEDQC}, and each communication round subjects idle qubits to decoherence for the duration of one inter-QPU block. Communication can therefore dominate a layer's noise cost, making the number of rounds the key scheduling quantity. The baseline assumes equal inter- and intra-QPU gate durations and noise parameters (Methods); Fig.~\ref{fig8} varies the inter-QPU cost.

With one communication qubit per QPU (Methods), NN communication preserves a
constant per-layer two-qubit-block depth, whereas the all-to-all
communication depth grows approximately linearly with the number of QPUs
[Fig.~\ref{fig7}(b)].
The NN and all-to-all architectures are not resource matched (Methods), so their comparison reflects variational expressivity and communication overhead together rather than connectivity alone.
Fig.~\ref{fig7}(c) compares noisy-VQE results as a function of the total layer
number $L$ for the brick-wall, distributed NN, and distributed all-to-all ansatzes.
At $L=1$ and $2$ the all-to-all circuit has the lowest mean final relative energy error, whereas for the tested $L\ge3$ the brick-wall has the lowest mean under this noise model. From $L=3$ to $4$, the all-to-all mean increases from $0.183(12)$ to $0.194(9)$ and the NN mean decreases from $0.20(3)$ to $0.186(2)$, both by less than their standard deviations at $L=3$; the brick-wall mean is nearly unchanged. Because each additional layer adds communication rounds without a resolved change in these means, the remaining comparisons for the twisted model use $(L,K,\chi)=(3,2,8)$.

Fig.~\ref{fig7}(d)--(f) shows VQE convergence at $(L,K,\chi)=(3,2,8)$ for the baseline two-qubit-gate duration $t_{\mathrm{2q}}=50\,\mu$s and the uniform-cost reference $(r_t,r_p)=(1,1)$, where $r_t$ and $r_p$ denote the ratios of inter-QPU to intra-QPU two-qubit-gate duration and biased-Pauli error probability, respectively (Methods).
In the noiseless simulations, the all-to-all architecture achieves the lowest
energy error among the three ansatzes, reflecting its greater expressivity 
arising from additional inter-QPU blocks, trainable parameters, and connectivity. 
The distributed NN circuit also reaches a lower mean noiseless error than the
brick-wall (Table~\ref{tab:training_objective}), the difference being
smaller than either sample standard deviation.
Although the single-QPU brick-wall and distributed NN circuits contain the
same number of two-qubit blocks per layer, they realize different coupling
graphs; in particular, QPU partitioning changes the variational connectivity
and introduces an NN communication edge, $(3,6)$, that is itself an
interaction pair of the target Hamiltonian (Supplementary Information). 
Under unmitigated noisy VQE, the brick-wall again has the lowest mean error among these three circuits [Fig.~\ref{fig7}(c)--(f)]. At the uniform-cost baseline, ZNE-assisted training gives the all-to-all circuit the lowest mean extrapolated error, while the NN and brick-wall means differ by less than their standard deviations (Table~\ref{tab:training_objective}).

The benefit of distribution depends on whether the connectivity gained through inter-QPU links outweighs the communication overhead once the circuits are trained.
The added connectivity is fixed by the circuit, whereas Fig.~\ref{fig8}
varies the cost through two parameters. The common cost factor
$r=r_t=r_p\in\{1,2,5,10\}$ ranges from equal-cost links to links ten times
slower and noisier, and the two-qubit-gate duration, $t_{\mathrm{2q}}=50\,\mu$s
(baseline) or $5\,\mu$s ($t_{\mathrm{2q}}/T_2=5\times10^{-3}$ and
$5\times10^{-4}$ with $T_1/T_2$ fixed), sets the idle decoherence per
communication round even at $r=1$.
To examine this trade-off with a single communication round,
Fig.~\ref{fig8} introduces a two-QPU circuit, the link circuit, alongside the three-QPU NN and all-to-all architectures. The two-QPU circuit, with partition $\{0,\ldots,3\}\,|\,\{4,\ldots,8\}$, implements the Hamiltonian interaction pair $(3,6)$ through a single inter-QPU block, while retaining the same block count as the brick-wall architecture (Methods). The single-QPU brick-wall circuit, which requires no inter-QPU operations, serves as the $r$-independent reference.
The link $(3,6)$ was selected by a noiseless scan over the interaction
pairs that connect the two parts. The scan and the comparisons of
Fig.~\ref{fig8} use independent pretraining runs, so the reported errors
are not those of the selection runs; the brick-wall reference
involves no selection (Supplementary Information).
For the brick-wall and three-QPU circuits, the $r=1$ ensembles at
$t_{\mathrm{2q}}=50\,\mu$s are those of Fig.~\ref{fig7}(c)--(f).

At the baseline two-qubit-gate duration $t_{\mathrm{2q}}=50\,\mu$s [Fig.~\ref{fig8}(a),(c)], 
the link circuit is the only distributed layout with a lower mean unmitigated error than the brick-wall at $r=1$ (Table~\ref{tab:training_objective}). The link circuit also has a lower mean ZNE-assisted error than the brick-wall at $r=1$ (Table~\ref{tab:training_objective}); at $r=2$ it has a higher one, $0.0271(4)$ versus $0.0264(21)$.
The three-QPU circuits require two (NN) or three (all-to-all) communication rounds per layer, increasing idle exposure and yielding larger unmitigated errors than the brick-wall reference across all $r$. Their ZNE-assisted errors exceed that of the brick-wall at $r=2$, $5$, and $10$.

Reducing the two-qubit-gate duration to $5\,\mu$s shortens the idle
intervals of communication tenfold and shifts the balance toward
connectivity [Fig.~\ref{fig8}(b),(d)].
At $r=1$, the link and all-to-all circuits have lower mean unmitigated errors than the brick-wall; the NN mean is slightly higher, $0.043(9)$ versus $0.0412(17)$. After ZNE-assisted training, all three means are lower than the brick-wall mean; the NN mean is $0.012(9)$ versus $0.0185(13)$. Both NN--brick-wall differences are smaller than the NN standard deviation.
The link circuit retains a lower mean unmitigated error at $r=2$. With ZNE-assisted training, the link and all-to-all means remain below the brick-wall mean at every tested $r$, whereas the NN mean at $r=10$, $0.020(9)$, exceeds it by less than the NN standard deviation (Supplementary Information).

Further calculations probe the role of the link. In a repeated noiseless scan of nine candidate links---the seven interaction pairs that connect the two parts, the chain bond $(3,4)$, and the non-interacting pair $(1,6)$---the link $(3,6)$ gives the lowest mean final error, followed by $(1,8)$, whereas $(3,5)$, $(0,8)$, and $(1,6)$ are worse than the brick-wall reference.
Replacing the link $(3,6)$ by the chain bond $(3,4)$, which leaves the
partition, block count, and depth unchanged but restores the coupling graph
of the brick-wall, raises the mean unmitigated error above that of the brick-wall at both durations. It also raises the mitigated error above that of the brick-wall at $50\,\mu$s, but not at $5\,\mu$s, where the mitigated error is $0.0158(22)$.
The nearest-neighbor Heisenberg reference (Methods) is the twisted model in
the qubit order of the chain, with the same exact energy, so that all its
interactions are between adjacent qubits. For this reference, all distributed
architectures have higher unmitigated energy errors than the brick-wall
circuit, with the difference increasing with communication depth
(Supplementary Information).
This reference also measures the cost of the fixed order: on the reordered chain, the
brick-wall circuit reaches a lower unmitigated error than every circuit in
the fixed order, including the link circuit, $0.0502(3)$ versus $0.133(8)$ at
$50\,\mu$s and $r=1$.
For the qubit order fixed by the MPS, these calculations are consistent with a benefit from links that realize Hamiltonian interactions otherwise long-range in the circuit, at least without mitigation. The mitigated chain-bond result at $5\,\mu$s shows that such a link is not required for every improvement.

A further reference makes the connectivity gain independent of the qubit order: for a nine-site Heisenberg model on a tree, with three arms meeting at a central site, no qubit order lets a chain cover all eight bonds.  A three-QPU circuit
that covers all bonds nevertheless has higher mean unmitigated errors than
single-QPU chain circuits with the same block count at every tested setting, and without
noise it is not more accurate than the chain with the same qubit order
(Supplementary Information).
Covering all bonds therefore does not by itself lower the noisy energy.

Taken together, Figs.~\ref{fig7} and \ref{fig8} show that a distributed ansatz
can retain a constant per-layer depth for $M\ge3$ when inter-QPU
communication is restricted to nearest-neighbor links [Fig.~\ref{fig7}(b)], and
that distributed circuits whose links realize long-range interactions of
the target Hamiltonian have lower mean unmitigated errors than the
single-processor brick-wall only when inter-QPU gates are at most twice as
slow and noisy as intra-QPU gates. With ZNE-assisted training at short gate
durations, they can remain more accurate even for inter-QPU gates ten times
as slow and noisy.
The simulations cover two and three QPUs. For larger $M$ the constant depth
of the NN schedule follows from the scheduling count of Fig.~\ref{fig7}(b)
and is not tested here.

\subsection{From MPS pretraining to noise-aware VQE}\label{sec:training_objective}

\begin{figure}[!htbp]
\centering
\includegraphics[width=0.85\textwidth]{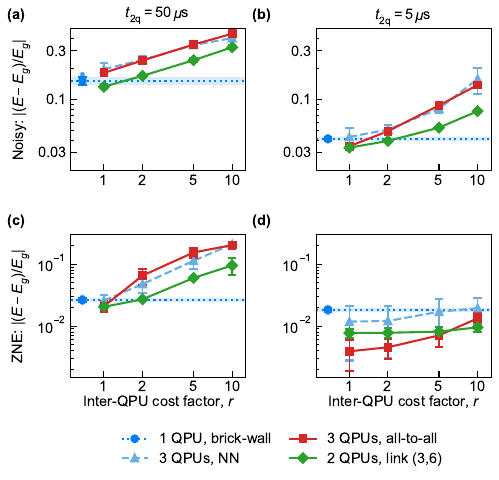}
\caption{Distributed quantum circuit performance versus the inter-QPU communication cost.
Final relative energy error after $10{,}000$ iterations for
$(L,K,\chi)=(3,2,8)$ as a function of the common inter-QPU cost factor
$r=r_t=r_p$, which multiplies the duration and the biased-Pauli error
probability of inter-QPU two-qubit operations relative to intra-QPU ones
(Methods). Panels (a),(c) use the baseline two-qubit-gate duration
$t_{\mathrm{2q}}=50\,\mu$s and (b),(d) $t_{\mathrm{2q}}=5\,\mu$s, i.e.\ $t_{\mathrm{2q}}/T_2=5\times10^{-3}$
and $5\times10^{-4}$ with $T_1/T_2$ fixed; all other noise parameters are
identical to those of Fig.~\ref{fig7}.
(a),(b) Unmitigated noisy VQE results; (c),(d) ZNE-assisted VQE results. The series correspond to the three-QPU NN and all-to-all circuits in Fig.~\ref{fig7} and a two-QPU circuit with partition $\{0,\ldots,3\}\,|\,\{4,\ldots,8\}$ and a single inter-QPU block on the Hamiltonian pair $(3,6)$. Markers and error bars indicate the mean and sample standard deviation over five runs initialized by independent MPS pretraining runs; each initial parameter set is reused at all values of $r$ (Methods). For the single-QPU brick-wall reference, which contains no inter-QPU operations, the five individual runs are shown as small circles to the left of $r=1$, while the dotted line and shaded band indicate the corresponding mean and standard deviation across each panel.
}
\label{fig8}
\end{figure}

Table~\ref{tab:training_objective} compares, for the circuits of
Fig.~\ref{fig8}, the pretrained start with the result of subsequent VQE under
noiseless, noisy, and ZNE evaluation, together with the corresponding state
and extrapolation diagnostics.
For each of the five independent pretraining realizations, the noiseless,
noisy, and ZNE-assisted runs at all noise settings share the initial parameters
of the MPS-pretrained layers and the appended near-identity layer (Methods).  This pretrained initial circuit is itself
a state-preparation circuit of the kind that is executed with fixed,
classically determined parameters
\cite{WeiMalz2026ParallelSequential,Jaderberg2026NormalMPSPreparation,Pennington2026SPT100Qubit,Scheer2025RG80MPS},
whereas VQE further optimizes the parameters using device measurements,
including after classical pre-optimization \cite{Khan2023PreoptimizingVQE}.
At fixed parameters $\boldsymbol{\theta}$ we define the noise-induced shift
$\Delta_{\mathrm{noise}}(\boldsymbol{\theta})\equiv
E_{\lambda=1}(\boldsymbol{\theta})-E_0(\boldsymbol{\theta})$, where
$E_{\lambda=1}$ is the noisy energy without folding (Methods) and $E_0$ the
noiseless energy; its values are reported in units of $|E_g|$. The noiseless energies at the noisy- and ZNE-trained
parameters serve as diagnostics of the states reached; the latter is denoted
ZNE$^{\dagger}$ in Table~\ref{tab:training_objective}.

At the pretrained start, the link and all-to-all circuits have markedly
lower mean noiseless errors than the brick-wall but, at $t_{\mathrm{2q}}=50\,\mu$s
and $r=1$, larger noise-induced shifts (Table~\ref{tab:training_objective}). Noiseless VQE leaves both circuits more accurate
than the brick-wall, with mean noiseless errors lower by factors of $2.5$ and
$5.6$. After noisy training, their advantage is smaller: the mean noiseless
errors at the trained parameters are lower by factors of $1.7$ and $1.1$, and
for all three circuits these errors exceed those after noiseless VQE. The reduction
of the noise-induced shift accounts for $53\%$ and $76\%$ of the mean
decrease in noisy error for the link and all-to-all circuits, respectively,
compared with $15\%$ for the brick-wall (Supplementary Information).
For the link circuit, the net effect depends on the communication cost: its
mean noisy error is below that of the brick-wall at the pretrained start for
$r\le2$, but after training only at $r=1$ [Fig.~\ref{fig8}(a); Supplementary
Information]. At $r=2$, the difference in noiseless error shrinks from $0.094$
at the pretrained start to $0.011$ after training, whereas the noise-induced
shift remains larger by $0.029$.
The tree reference of Sec.~\ref{sec:distributed_qpu}, whose three-QPU circuit
starts from a lower noiseless error than the chain circuits (Supplementary
Information), likewise shows that a lower noiseless error at the pretrained
start need not yield a lower noisy error after training. At $50\,\mu$s, unlike
for the chain circuits, the noiseless error of the tree circuit increases during
noisy training.
For the distributed circuits, MPS pretraining accelerates the initial decrease
in energy error but does not lower the mean final unmitigated error in these
runs: at $50\,\mu$s
and $r=1$, runs from random initial parameters reach similar or lower mean
final unmitigated errors, whereas for the brick-wall they reach a higher one
(Supplementary Information).

The ZNE estimate can be written as
$E_{\mathrm{ZNE}}(\boldsymbol{\theta})=E_0(\boldsymbol{\theta})+|E_g|\,b(\boldsymbol{\theta})$,
where the added term $|E_g|\,b$, with $b$ the normalized extrapolation error,
is a systematic error of the estimator rather than a change in the energy of
the executed circuit.  At $50\,\mu$s and $r=1$, ZNE-assisted training lowers
$b$ together with the noiseless error; for the all-to-all circuit, $b$
decreases from $0.051$ at the pretrained start to $0.009$
(Table~\ref{tab:training_objective}).  For the tree circuit at $50\,\mu$s, however, the extrapolated energy
decreases between iterations $1{,}000$ and $10{,}000$ while the noiseless
energy at the same parameters increases, and the extrapolated energies end
below $E_g$.
At $5\,\mu$s and $r\le2$, $b$ is slightly negative after ZNE-assisted
training in most runs of Fig.~\ref{fig8} (Supplementary Information).  A lower extrapolated energy therefore
need not indicate a lower noiseless energy of the optimized circuit, as for
the ladder in Fig.~\ref{fig6}(a).
Moreover, applying the extrapolation once at the final parameters of
noisy VQE does not reach the energies of ZNE-assisted training: at
$50\,\mu$s and $r=1$ the extrapolated error at the noisy-trained parameters
is higher than after ZNE-assisted training for every circuit
(Table~\ref{tab:training_objective}), and at $5\,\mu$s it nearly equals the
noiseless error of the noisy-trained parameters (Supplementary Information).

\begin{table}[t]
\caption{MPS-pretrained start and subsequent VQE for the circuits of
Fig.~\ref{fig8} at $t_{\mathrm{2q}}=50\,\mu$s and $r=1$.  The pretrained initial
circuit consists of the MPS-pretrained layers and the appended near-identity
layer;
within each circuit, all methods share the same five initial parameter sets.
After $10{,}000$ iterations, noiseless, noisy, and ZNE-assisted VQE are
evaluated with the noiseless, noisy, and extrapolated energies, respectively;
an asterisk marks the noiseless relative energy error at the noisy-trained
parameters and a dagger the corresponding diagnostic at the ZNE-trained
parameters (ZNE$^{\dagger}$).  For noisy VQE, the ZNE column gives the extrapolated error evaluated once at the noisy-trained parameters.  Values are mean relative energy errors over five runs with
the sample standard deviation in the last digits.}
\label{tab:training_objective}
\small
\setlength{\tabcolsep}{4pt}
\renewcommand{\arraystretch}{1.16}
\begin{tabular}{@{}lccc@{}}
\toprule
Stage & Noiseless & Noisy & ZNE \\
\midrule
\multicolumn{4}{@{}l}{\textit{1 QPU, brick-wall}} \\
Pretrained start & $0.169(17)$ & $0.303(12)$ & $0.178(17)$ \\
Noiseless VQE & $0.020(6)$ & -- & -- \\
Noisy VQE & $0.041(26)^{*}$ & $0.153(13)$ & $0.048(24)$ \\
ZNE-assisted VQE & $0.0205(19)^{\dagger}$ & -- & $0.0264(21)$ \\
\addlinespace[2pt]
\multicolumn{4}{@{}l}{\textit{2 QPUs, link $(3,6)$}} \\
Pretrained start & $0.0756(4)$ & $0.2420(15)$ & $0.0907(5)$ \\
Noiseless VQE & $0.0081(15)$ & -- & -- \\
Noisy VQE & $0.024(3)^{*}$ & $0.133(8)$ & $0.034(4)$ \\
ZNE-assisted VQE & $0.0155(11)^{\dagger}$ & -- & $0.0205(18)$ \\
\addlinespace[2pt]
\multicolumn{4}{@{}l}{\textit{3 QPUs, NN}} \\
Pretrained start & $0.16(3)$ & $0.375(29)$ & $0.19(3)$ \\
Noiseless VQE & $0.017(10)$ & -- & -- \\
Noisy VQE & $0.08(6)^{*}$ & $0.20(3)$ & $0.10(6)$ \\
ZNE-assisted VQE & $0.017(8)^{\dagger}$ & -- & $0.026(7)$ \\
\addlinespace[2pt]
\multicolumn{4}{@{}l}{\textit{3 QPUs, all-to-all}} \\
Pretrained start & $0.072(7)$ & $0.333(16)$ & $0.123(10)$ \\
Noiseless VQE & $0.0036(5)$ & -- & -- \\
Noisy VQE & $0.037(5)^{*}$ & $0.183(12)$ & $0.059(9)$ \\
ZNE-assisted VQE & $0.012(3)^{\dagger}$ & -- & $0.021(4)$ \\
\botrule
\end{tabular}
\end{table}

\section{Discussion}

Our central results are that circuit realizations derived from the same
target MPS family, with matched per-layer two-qubit-block resources, can
show comparable ideal VQE and gradient-scale behavior yet different
performance under noise, and that, at the baseline gate duration, training
under noise narrows the noiseless advantage that the link and all-to-all
circuits hold at the pretrained start.
Under the combined gate-local noise model at the baseline two-qubit-gate
duration, the brick-wall layout, which has the shallowest realized schedule, gives the
lowest unmitigated error among the three single-processor layouts and, after
ZNE-assisted training, the lowest noiseless energy error.
Because all three layouts contain the same number of two-qubit blocks per
layer, this separation reflects how those blocks are parallelized and
scheduled rather than their nominal count, consistent with hardware
observations that shallow MPS state-preparation circuits outperform deeper staircase
circuits despite larger CNOT counts \cite{Jaderberg2026NormalMPSPreparation}.
Finite gate duration and idle-time
exposure are therefore practical circuit resources alongside gate count and parameter number.
The gate-error-dominant control of Sec.~\ref{sec:noisy_results} supports
this reading: at equal gate counts the noise-induced shifts of the three
layouts then no longer favor the brick-wall.

The same scheduling difference shapes error mitigation.
Global unitary folding repeats each architecture's gate sequence, so the deeper ladder and mixed-canonical schedules accumulate more idle-time decoherence during folded evaluations than the shallow brick-wall schedule.
ZNE therefore does not remove architecture-dependent execution costs, and its
effectiveness remains tied to the baseline circuit architecture \cite{Temme2017ZNE,Endo2018QEM}.
For the ladder, the extrapolated energy lies below the noiseless energy at
the final parameters, so part of its mitigated improvement is extrapolation
error rather than a better state (Sec.~\ref{sec:training_objective}).

The comparable ideal-circuit gradient scales provide an interpretive
control: within the gradient diagnostic examined here, the brick-wall ansatz
shows no gradient advantage that could account for its noise robustness,
consistent with the role of local depth in gradient concentration \cite{Zhang2024AbsenceBarrenPlateaus}.
This diagnostic does not, however, characterize the full optimization landscape, where local minima and traps remain relevant \cite{Bittel2021TrainingVQANPHard,YouWu2021ExponentiallyManyLocalMinima,AnschuetzKiani2022SwampedWithTraps,Anschuetz2025UnifiedTheoryQNNLossLandscapes}, and landscape results for MPS manifolds need not extend to multilayer circuits initialized with MPS-derived parameters \cite{ZhangXiang2026AbsencePoorLocalMinimaMPS}.

The ideal encoding and noisy-VQE results together illustrate a practical
fidelity--depth trade-off in MPS-to-circuit encoding
\cite{BenDov2024ApproximateEncoding,Iaconis2024NormalDistributionsMPS}.
Additional encoding layers improve the representation of the target MPS
(Fig.~\ref{fig4}), but, when they increase the total layer number, they also
lengthen the circuit executed at each subsequent VQE iteration.
Maximizing encoding fidelity therefore need not yield the most
useful hardware-executable initialization.
In contrast to shallow brick-wall state-preparation circuits compiled from DMRG states and executed directly on superconducting hardware \cite{Pennington2026SPT100Qubit}, MPS-pretrained layers form only the initial part of a repeatedly evaluated VQE circuit, so that their quality must be assessed together with execution and optimization costs.

The distributed results extend this picture to a trade-off between the
connectivity gained through inter-QPU links and the idle-time decoherence 
incurred during communication rounds.
In the circuits studied, the distributed circuits that reach lower mean
unmitigated errors than the single-QPU brick-wall circuit at equal gate
quality ($r=1$) are among those whose links realize couplings that the chain lacks
in the qubit order fixed by the MPS; each communication round they add
exposes the idle qubits to decoherence for the duration of one inter-QPU
block (three decomposed two-qubit gates; Methods).
In Fig.~\ref{fig8}, the link circuit, with one Hamiltonian-matched link
and one communication round, is the only distributed layout below the
brick-wall without mitigation at the baseline duration and $r=1$, whereas
the three-QPU circuits, with two or three communication rounds, reach or
fall below it only after ZNE at $r=1$.
Without a connectivity gain, as for the nearest-neighbor Heisenberg chain
and the chain-bond link of Sec.~\ref{sec:distributed_qpu}, distribution
mainly adds communication overhead. For the distributed circuits of
Fig.~\ref{fig8}, the benefit of distribution at the baseline duration
decreases monotonically with $r$.

Whether a connectivity gain is available to a distributed circuit depends on the
interaction graph: a suitable qubit order lets a chain of nearest-neighbor blocks cover a
path-like graph completely, but not a graph with branch points or cycles.
A modular architecture with one communication qubit per QPU can recover the
couplings at a branch point by placing that vertex on a communication qubit.
The tree calculation shows, however, that
covering all bonds does not by itself lower the noisy energy, and in the
twisted model only a subset of the interaction pairs improves the final
noiseless VQE error when implemented as a link (Supplementary Information).
An effective distributed architecture should therefore target beneficial
missing interactions with minimal communication depth, rather than maximize
connectivity alone.
In this respect, the NN QPU-path schedule carries the brick-wall depth
advantage to modular processors: additional QPUs expand the accessible
register without increasing the per-layer two-qubit-block depth, whereas the
all-to-all communication depth grows with the number of QPUs.

Mitigation does not remove the communication cost: at the baseline two-qubit-gate
duration, the link and all-to-all circuits have lower mean ZNE-assisted
errors than the brick-wall only at $r=1$, and the NN circuit a comparable
one, whereas at $5\,\mu$s the link and all-to-all circuits do so at every
tested $r$. At the baseline duration, the noiseless energy at the final ZNE
parameters of the link circuit nevertheless remains below the brick-wall value up to $r=5$, although its
extrapolated error is higher at $r=2$ and $5$.
The deviation of the extrapolated energy from the noiseless energy of the
optimized circuit also grows with the communication cost, on average from
below $10^{-3}$ at $r\le2$ to $5\times10^{-3}$ for $r\ge5$ at
$t_{\mathrm{2q}}=5\,\mu$s, and at $50\,\mu$s from $0.005$--$0.009$ at $r=1$ to
$0.06$ at higher $r$ (Supplementary Information).  
Thus, increasing communication cost limits not only direct noisy execution 
but also the accuracy of extrapolation-based error mitigation.

Comparing the pretrained and trained circuits shows how noisy training
changes the balance between noiseless accuracy and the noise-induced shift.
At $50\,\mu$s and $r=1$, noisy training narrows the noiseless advantage
that the link and all-to-all circuits hold at the pretrained start and after
noiseless VQE, while also reducing their noise-induced shifts. A lower noisy error then requires the
remaining noiseless advantage to outweigh an excess shift that, for
the link circuit, increases with the inter-QPU cost. The benefit of
distribution should therefore be judged on circuits trained under the
noise of the intended device rather than on the pretrained initial
circuits or those obtained by noiseless VQE.
A single extrapolation after noisy VQE does not replace ZNE-assisted
training: it estimates the noiseless energy of the state reached under
noise, with a residual bias at the baseline duration, and remains above the
ZNE-assisted result (Table~\ref{tab:training_objective}).
For the distributed circuits the pretraining mainly accelerates the initial
decrease in energy error, whereas random
initialization reaches similar or lower mean final unmitigated errors at
$t_{\mathrm{2q}}=50\,\mu$s and $r=1$ (Supplementary Information).

More broadly, these findings suggest that TN pretraining should be viewed as
a hardware-aware co-design problem rather than an isolated state-encoding step.
Whether a classically optimized tensor-network state retains its
advantage in an executable circuit is decided jointly by the encoding
fidelity, the scheduled gate duration and idle exposure, the QPU partitioning
and communication topology, and the training method and error-mitigation
strategy.
Pretraining strategies could therefore weigh encoding fidelity against
scheduled execution cost, and hybrid TN--quantum workflows that retain part
of the variational representation on a classical processor offer a
complementary route to controlling execution depth
\cite{Huang2023TN-VQA,Schuhmacher2025HybridTTN}.
The noisy-VQE comparisons use eight or nine qubits, at most three QPUs, and
a phenomenological noise model without measurement shot noise.
How the relative contributions of MPS-informed initialization and
architecture-dependent scheduling evolve with system size, noise regime, and
target Hamiltonian remains to be examined.

\section{Methods}\label{sec:methods}

This section describes the circuit constructions, noise model, optimization
protocol, error-mitigation procedure, distributed-QPU architecture, and
statistical conventions used in the numerical simulations.
Further derivations and numerical details are provided in the Supplementary
Information.

\subsection{MPS-to-circuit encoding and depth-efficient circuit layouts}
\label{subsec:encoding}
In this work, MPS-to-circuit encoding denotes the optimization of $K$
circuit layers to approximate a classically optimized target MPS of bond
dimension $\chi$.
MPS pretraining refers to using the resulting parameters to initialize
the first $K$ layers of a total $L$-layer VQE circuit, with all circuit
parameters remaining trainable during VQE. This procedure provides a
structured starting point that captures the relevant low-entanglement
correlations.
Sequential and brick-wall matrix-product-disentangler constructions
\cite{Ran2020,Rudolph2024MPSDecomposition,Szoldra2026ScalableMPSPreparation}
and variational MPS compilation
\cite{BenDov2024ApproximateEncoding,MansurogluSchuch2026VariationalDisentanglement,Jaderberg2026NormalMPSPreparation}
have been extensively discussed as MPS-to-circuit encoding strategies.
Direct TN simulation of parameterized circuits provides another route to
classical pretraining \cite{Khan2023PreoptimizingVQE}.
We focus here on the use of MPS-to-circuit encoding for VQE initialization
\cite{Dborin2022MPSPretraining,Rudolph2023SynergisticPretraining} and on the
circuit-architecture choices considered in this study.
The Supplementary Information provides explicit derivations of the encoding procedures, including the exact $\chi=2$ construction and the iterative decomposition protocol for higher bond dimensions. 

We consider three representative circuit layouts with varying depths---ladder, mixed-canonical, and brick-wall---illustrated in Fig.~\ref{fig2}. 
For a given layer, all three ansatzes employ the same number of two-qubit 
unitary blocks, $N-1$, but differ in how these blocks are scheduled 
in parallel. 
Circuit depth is measured in units of two-qubit unitary blocks, 
assuming that operations acting on disjoint qubit pairs are executed simultaneously. 
For resource estimates, each two-qubit unitary block is treated as
an abstract generic $\mathrm{SU}(4)$ operation. In the noisy
simulations, we use a KAK-based decomposition of each block into
local single-qubit rotations and parameterized IsingXX, IsingYY,
and IsingZZ interactions \cite{Vatan2004OptimalTwoQubits,Vidal2004ThreeCNOTs}. 
We note that these decomposed operations define the gate primitives of 
our simulation and are not assumed to coincide with the native gate set of any 
particular hardware platform.
Table~\ref{tab:ansatz_summary} summarizes the resulting circuit depths 
and MPS encoding fidelities for each ansatz.

The ladder and mixed-canonical layouts both admit exact $\chi=2$ encodings, 
with per-layer depths of $N-1$ and $\lceil N/2\rceil$, respectively. 
By contrast, the brick-wall layout consists of alternating odd and even sublayers, 
yielding a constant per-layer depth of $2$, independent of $N$. 
This parallel structure does not, in general, support an exact $\chi=2$ encoding. 
The limitation is structural: 
a single brick-wall layer has a finite correlation range and
therefore cannot exactly represent a generic $\chi=2$ MPS
\cite{Lin2021CompressedCircuits}, whereas exact sequential constructions
propagate a virtual ancilla of dimension $\chi$ along the chain
\cite{Ran2020,Schon2005ExactEncoding}.
Such sequential ordering is naturally compatible 
with the ladder and mixed-canonical layouts,  
but not with the fixed odd-even parallel scheduling of a single brick-wall layer. 
Despite this restriction, the fidelity gap in Table~\ref{tab:ansatz_summary}
is small, indicating that the target states considered here lie 
close to the corresponding brick-wall circuit manifold 
even without a generic exact mapping. 
For larger $\chi$, the primary consideration is 
how efficiently an MPS can be approximated 
using a limited number of layers. 
Although the brick-wall ansatz yields lower state fidelity than 
the ladder and mixed-canonical layouts at $\chi=8$ with $K=1$, 
the three become nearly indistinguishable at $N=8$ as $K$ increases, demonstrating
that at this system size a shallow multilayer approximation compensates
for the absence of an exact single-layer brick-wall encoding.

\subsection{Noise models} \label{sec:noise_model}
To evaluate the robustness of different MPS-pretrained PQC ansatzes, we incorporate two representative noise channels:
\emph{(biased) Pauli errors} and \emph{thermal relaxation}.  
Noise is applied at the decomposed-gate level, following each
single-qubit rotation and each IsingXX, IsingYY, or IsingZZ
operation generated by the generic $\mathrm{SU}(4)$ block
decomposition.

A \emph{(biased) Pauli channel} models stochastic Pauli errors as
\begin{equation}
\mathbb{E}_{P}(\rho)=(1-p)\rho+\sum_{k\in\{X,Y,Z\}}p_k\,\sigma_k\rho\sigma_k ,
\end{equation}
where $p=p_X+p_Y+p_Z$ is the total error probability \cite{NielsenChuang, QiskitAerThermalRelaxationError}.
We adopt a phase-flip-biased error model with $p_Z=0.998p$ and $p_X=p_Y=0.001p$.
For a single-qubit gate, $p=p_s=10^{-5}$. 
For a two-qubit gate, independent copies of the same single-qubit biased Pauli
channel act on the two active qubits with $p=p_d=10^{-4}$ per active
qubit; thus $p_d$ is not the total two-qubit-block error probability.

To capture decoherence accumulated during circuit execution, we apply a single-qubit
thermal-relaxation channel, $\mathbb{E}_{T}$, after every
gate and throughout each idle interval of duration $t_g$. 
This channel is parameterized by the energy-relaxation and 
dephasing times, $T_1$ and $T_2$ \cite{QiskitAerThermalRelaxationError, IBMQuantumBuildNoiseModels, BravoMontes2024EPJQT}.
Populations relax toward the ground state (excited-state population $p_e=0$) with probability
$p_{\mathrm{reset}}=1-\exp(-t_g/T_1)$, 
while coherences decay by a factor $\exp(-t_g/T_2)$.
Throughout this work, we use $T_1=500\,\mathrm{ms}$ and $T_2=10\,\mathrm{ms}$,
with baseline single- and two-qubit-gate durations of
$t_{\mathrm{1q}}=10\,\mathrm{ns}$ and
$t_{\mathrm{2q}}=50{,}000\,\mathrm{ns}=50\,\mu$s,
respectively. 
These parameters are not intended to reproduce a calibration of any
specific hardware platform, as native gate sets and characteristic
operation and coherence times vary across quantum-computing platforms.
Rather, they define a phenomenological finite-duration noise baseline in
which gate-local errors coexist with relaxation and dephasing during active
and idle periods.
The ordering of the gate-error and thermal-relaxation channels is the
same as in the device noise models constructed by Qiskit Aer
\cite{IBMQuantumBuildNoiseModels}, and the thermal component is the same
zero-temperature relaxation channel as in Qiskit Aer for $T_2\le T_1$, the
regime used here \cite{QiskitAerThermalRelaxationError}.  The present model
differs in that it uses independent, phase-flip-biased Pauli errors and uniform parameters,
treats Ising rotations as two-qubit gate primitives, and omits readout errors.

To place the resulting error scales in context, we compare them with four
calibration snapshots of IBM processors provided as fake backends in Qiskit
IBM Runtime \cite{QiskitIBMRuntimeFakeProvider}.  At
$t_{\mathrm{2q}}=50\,\mu$s, the two-qubit-gate infidelity
($4.2\times10^{-3}$) falls within the range of the four reported median
two-qubit-gate errors, and the idle infidelity over the same duration
($1.7\times10^{-3}$) within the range of the corresponding relaxation
estimates.  At $5\,\mu$s, both quantities are below all four
reference values, and the single-qubit-gate infidelity is below the median
calibrated SX-gate error of each snapshot (Supplementary Information).
Under this baseline, schedule-dependent finite-duration decoherence remains
appreciable; the sensitivity to the assigned two-qubit-gate duration is
examined separately in the Supplementary Information.

Idle qubits undergo thermal relaxation for the duration of the 
scheduled sublayer: $t_{\mathrm{1q}}$ for a single-qubit-only sublayer and
$t_{\mathrm{2q}}$ for sublayers containing at least one two-qubit gate. 
Folded inverse gates introduced for ZNE are assigned the same durations.
Fig.~\ref{fig1}(b) illustrates noise channels acting
after gates and during idle intervals, 
while Fig.~\ref{fig5}(a) summarizes the gate-level noise implementation. 
The Supplementary Information details the channel composition. 
All noisy simulations employ the combined gate-local noise model.
\subsection{VQE optimization and zero-noise extrapolation} 
\label{sec:vqe_description}

We benchmark the MPS-pretrained ansatzes using the VQE~\cite{Peruzzo2014VQE,Tilly2022VQEReview}. 
The target Hamiltonian is a Pauli-normalized one-dimensional Heisenberg
model with a ``twisted'' pairing pattern, the open-boundary chain in a
permuted site order,
\begin{equation}
\label{eq:twisted_hamiltonian}
H =
\sum_{i=0}^{\lfloor N/2 \rfloor -1}
\boldsymbol{\sigma}_i \cdot \boldsymbol{\sigma}_{N-1-i}
+
\sum_{i=1}^{\lfloor (N-1)/2 \rfloor}
\boldsymbol{\sigma}_i \cdot \boldsymbol{\sigma}_{N-i},
\end{equation}
where $\lfloor \cdot \rfloor$ denotes the floor function and
$\boldsymbol{\sigma}_i=(\sigma_i^x,\sigma_i^y,\sigma_i^z)$ is the vector of
Pauli operators acting on site $i$.  
This Pauli-normalized Hamiltonian is four times the conventional 
spin-operator form expressed in terms of
$\mathbf{S}_i=\boldsymbol{\sigma}_i/2$. 
The two conventions share the same eigenstates and differ 
only by an overall energy scale. 
We adopt this Hamiltonian because the permutation preserves the spectrum
of the conventional open-boundary chain while inducing a nontrivial
interaction pattern in the original qubit ordering. 
This permutation increases the circuit complexity without altering the underlying physics.
Conversely, correlation-informed qubit permutations have been used in VQE to
reduce the required circuit depth~\cite{Tkachenko2021PermVQE}.
The reference energy $E_g$ used to compute the relative energy error is
obtained by exact diagonalization of the finite-size Pauli-normalized
Hamiltonian for the corresponding system size $N$.

For each architecture, we minimize the variational energy $E(\boldsymbol{\theta})$
using the Adam optimizer~\cite{Kingma2014Adam} with an initial learning rate of
$10^{-2}$ and the gradient norm clipped at $1$; the learning rate is halved,
down to a minimum of $10^{-4}$, after every $30$ consecutive iterations without
an improvement of more than $10^{-12}$ in the best objective value.
The illustrative noiseless trajectory in Fig.~\ref{fig1}(b) is obtained with L-BFGS.
The optimization is performed for $2\times 10^4$ iterations in the $N=8$ ideal-circuit simulations and for $10^4$ iterations in the noisy and ZNE-assisted simulations and in all $N=9$ simulations.  The energies of the pretrained initial circuit are evaluated before VQE optimization. 
All energies are exact expectation values from state-vector or density-matrix simulation, without measurement shot noise. 
Each circuit comprises $L=K+e$ layers of two-qubit unitary blocks, 
where the first $K$ layers are initialized by MPS pretraining and the remaining $e$ layers are added to increase the circuit's expressive power. 
The additional (non-pretrained) layers are initialized by independently sampling all 
gate parameters from the normal distribution 
$\mathcal{N}(0,(10^{-2})^2)$.
This initialization produces near-identity gates, consistent with 
previous MPS-pretraining studies~\cite{Rudolph2023SynergisticPretraining}. 
At a fixed total layer count $L$, we vary the number of pretrained layers $K$ and 
the MPS bond dimension $\chi$ to evaluate how MPS pretraining influences ideal-circuit performance across
different circuit structures.
For the gradient-variance diagnostics in Fig.~\ref{fig3}(d),(e), 
we compute the unbiased sample variance over $100$ random parameter realizations,
each drawing every two-qubit block as an independent Haar-random
$\mathrm{SU}(4)$ unitary without MPS pretraining,
for the \texttt{RYY} (IsingYY) parameter in the first two-qubit unitary block 
according to each ansatz's construction order; for the mixed-canonical ansatz, 
this corresponds to the central two-qubit block.  
The selected parameter is matched by construction order rather than 
by physical bond across the three architectures.

To mitigate noise, we employ zero-noise extrapolation (ZNE)~\cite{Temme2017ZNE,LiBenjamin2017EfficientDistributed,Endo2018QEM}. 
Noise scaling is implemented by global unitary folding~\cite{GiurgicaTiron2020DigitalZNE} 
with fold factors $\lambda=1,3,5$.
The zero-noise expectation value is then estimated by quadratic extrapolation 
in $\lambda$, using the three-point Richardson interpolant defined
by the folded values at $\lambda=1,3,5$ and evaluated at $\lambda=0$.
The resulting estimate is given by
\begin{equation}
\langle O\rangle_{0} \approx \frac{15}{8}\langle O\rangle_{1}
-\frac{5}{4}\langle O\rangle_{3}
+\frac{3}{8}\langle O\rangle_{5}.
\end{equation}
All ZNE simulations use deterministic noisy expectation values without shot
sampling; consequently, finite-shot statistical fluctuations are absent.
During ZNE-assisted VQE, the Richardson-extrapolated loss is constructed 
at each VQE iteration from matched fold-1, fold-3, and fold-5 circuit evaluations
and used directly as the optimization objective. 
Because the Richardson combination extrapolates beyond the sampled fold
factors with a negative coefficient, the extrapolated estimator is neither
variational nor guaranteed to improve every optimization trajectory.
For the noisy-VQE runs of Fig.~\ref{fig8} at $r=1$, the same estimator
is also evaluated once at the final parameters, from matched fold-1,
fold-3, and fold-5 evaluations without further optimization
(Table~\ref{tab:training_objective}).
Further details of the folding protocol and zero-noise extrapolation
are provided in the Supplementary Information.

\subsection{Distributed-QPU architecture}
\label{sec:distributed_methods}

The distributed ansatz extends the odd-even brick-wall pattern to a modular
architecture of $M$ interconnected QPUs.
We assume that each QPU contains a designated communication qubit for
inter-QPU entangling operations, whereas the remaining qubits are used
exclusively for local circuit operations.
This restriction is analogous to the communication-qubit bottleneck
considered in compilers for distributed quantum computing
\cite{Ferrari2021CompilerDQC}, in which reserved communication qubits,
distinct from the data qubits, store the Bell pairs consumed by each remote
gate; here, the communication qubit is one of the qubits of the chain and also
participates in the local blocks, and inter-QPU operations are modeled as abstract
two-qubit unitary blocks rather than through an explicit gate-teleportation protocol.
In every layer, intra-QPU two-qubit unitary blocks are first arranged in a
local brick-wall pattern within each QPU; this local stage is followed by
inter-QPU two-qubit blocks acting on the designated communication qubits,
scheduled in a staggered pattern that maximizes parallel operations across
the communication qubits.
We consider two communication topologies, nearest-neighbor (NN) and
all-to-all, as illustrated in Fig.~\ref{fig7}(a).
In the NN architecture, inter-QPU operations are restricted to adjacent
QPUs, allowing each layer to retain the same number of two-qubit unitary
blocks as the brick-wall baseline while preserving a constant per-layer
block depth under the single-communication-qubit path model, in which the
QPUs form a path and each contributes one communication qubit.
In the all-to-all architecture, inter-QPU two-qubit blocks act between every
pair of QPUs; the number of communication links grows quadratically with the
number of QPUs, the required communication rounds can no longer remain
constant, and the per-layer depth increases approximately linearly with the
number of QPUs [Fig.~\ref{fig7}(b)].
The NN and all-to-all distributed ansatzes are therefore not resource matched:
compared with the NN architecture, the all-to-all construction introduces
additional inter-QPU two-qubit blocks, trainable parameters, and communication
depth.
The resulting depth expressions and the block and parameter counts of each
architecture are derived in the Supplementary Information.

In the baseline distributed simulations, inter-QPU and intra-QPU two-qubit
gates are assigned identical durations and noise parameters, providing a
uniform-cost reference $(r_t,r_p)=(1,1)$ with
$r_t=t_{\mathrm{2q},\mathrm{inter}}/t_{\mathrm{2q},\mathrm{intra}}$ and
$r_p=p_{d,\mathrm{inter}}/p_{d,\mathrm{intra}}$; because communication
penalties would act only on the distributed circuits, this reference favors them
in the unmitigated architecture comparison.
The inter-QPU cost scan of Fig.~\ref{fig8} sets both ratios to a
common factor $r=r_t=r_p\in\{1,2,5,10\}$.
The factor $r_t$ scales the duration of every scheduled sublayer 
containing an inter-QPU gate and, consequently, the thermal-relaxation
exposure of the idle qubits within that sublayer. 
By contrast, $r_p$ scales the biased-Pauli error probability 
of the inter-QPU gate itself. 
We perform this scan at $t_{\mathrm{2q}}=50\,\mu$s and
$5\,\mu$s, with all other noise parameters held fixed.
The two-QPU link circuit of Fig.~\ref{fig8} partitions the chain into qubits $\{0,\ldots,3\}$ and
$\{4,\ldots,8\}$, applies the local brick-wall blocks within each part, and
adds one inter-QPU block per layer between the communication qubits $3$ and
$6$, so that its layer contains the same eight blocks as a brick-wall layer
but has depth $D_2=3$. The link $(3,6)$ was selected among the interaction pairs of
the Hamiltonian that connect the two parts by its final noiseless VQE error; it also has the lowest mean final error of the candidates in a repeated scan whose pretraining runs are independent of those of Fig.~\ref{fig8} (Supplementary Information).
The chain-bond control circuit keeps the same QPU partitioning but replaces the $(3,6)$ inter-QPU link with the chain bond $(3,4)$, thereby recovering the coupling graph of the single-QPU brick-wall architecture.
The nearest-neighbor Heisenberg reference uses the nine-qubit chain $H=\sum_{i=0}^{N-2}\boldsymbol{\sigma}_i\cdot\boldsymbol{\sigma}_{i+1}$ 
with a $\chi=8$ DMRG target MPS, while keeping the same circuit layouts. 
For this reference, the three-QPU circuits place the communication qubits on
qubits $2$, $3$, and $6$ instead of $0$, $3$, and $6$, allowing the cut bond
$(2,3)$ to be restored by an inter-QPU block. 
With only one communication qubit per QPU, however, 
the second cut bond $(5,6)$ cannot be simultaneously restored.
For the distributed simulations in Figs.~\ref{fig7} and \ref{fig8}, Table~\ref{tab:training_objective}, and the corresponding reference calculations, the local sweeps of the MPS pretraining use untruncated environments (bond dimension $16$, exact for $N=9$) and at most $500$ sweeps. Each of the five runs uses an independent MPS-pretraining realization; the resulting parameter sets are reused for the noiseless, noisy, and ZNE-assisted runs of the same circuit at every noise setting. By contrast, for each single-processor circuit configuration in Figs.~\ref{fig3}(a)--(c), \ref{fig5}, and \ref{fig6}, the noiseless, noisy, and ZNE-assisted ensembles use separate MPS pretrainings. Runs within each ensemble share one pretraining and differ only in the random initialization of the appended layers.
As a control, the three training methods were also run from the same
five stored initial-parameter sets for each single-processor circuit at
$(L,K,\chi)=(3,2,8)$ and $(4,3,8)$, with the pretraining settings of the
distributed calculations, and the $(3,2,8)$ circuits were additionally
optimized under noise with $p_d=5\times10^{-3}$ at
$t_{\mathrm{2q}}=5\,\mu$s (Supplementary Information).

\subsection{Statistical analyses}
Unless stated otherwise, each reported value is the mean over five runs with
the sample standard deviation given in parentheses in
the last digits; the $N=8$ noiseless references use $20$ runs.  In the distributed calculations, the noiseless, noisy, and ZNE-assisted runs
of a circuit at all its noise settings start from the same five pretrained
parameter sets, so the training methods are compared run by run and values at
different noise settings are not independent samples.  Different circuits do
not share initial states and are compared through their means and standard
deviations; no significance tests are performed.  Results obtained with
different pretraining protocols are not pooled (Supplementary Information).
The simulations use exact expectation values, and final energies are those
at the last iteration of each run.

\subsection{Use of large language models}
Large language model tools were used to assist with language editing,
manuscript consistency checks, and preparation of data and code availability
materials. They were not used to generate scientific claims, numerical results,
or simulation outputs, and all AI-assisted text was reviewed and validated by
the authors.

\backmatter

\section*{Data availability}
The source data underlying the figures and Tables~1 and 2 of this article will be made publicly available in a public repository upon publication. Other data are available from the corresponding authors upon reasonable request.

\section*{Code availability}
The scripts that reproduce the figures and tables from the released data will be deposited with the data. The simulation code is available from the corresponding authors upon reasonable request.

\section*{Acknowledgements}
This work was supported, in whole or in part, by the National Research Foundation of Korea (NRF), 
funded by the Korean government (MSIT) (Grant Nos. RS-2023-NR119925, 
RS-2023-NR119928, RS-2024-00413957, RS-2024-00442710, and RS-2025-25454922), 
the Institute for Information \& Communications Technology Planning \& Evaluation (IITP) 
(Grant Nos. RS-2025-02219034 and RS-2025-25464252), 
and the KIST Institutional Program (Grant No. 26E0001).

\section*{Author contributions}
S.H. and J.L. conceived the project. S.H. developed the numerical simulations and performed the data analysis. S.-S.B.L. and J.L. supervised the project and contributed to the interpretation of the results. All authors discussed the results and contributed to writing the manuscript.

\section*{Competing interests}
The authors declare no competing interests.

\bibliography{references}

\clearpage


\section*{Supplementary material}

# 1 MPS-to-circuit construction

This section fixes the tensor notation used for the MPS-pretrained ansatzes, summarizes the exact $\chi = 2$ ladder and mixed-canonical encodings, and states how the higher-bond-dimension circuits are constructed iteratively. The notation follows standard open-boundary MPS canonical forms and their use in sequential state preparation [1–4].

## 1.A Canonical MPS and exact $\chi = 2$ encoding

For an $N$-qubit open-boundary MPS,

$$|\psi_{\mathrm{MPS}}\rangle = \sum_{\{s_n\}} \sum_{\{\alpha_n\}} A^{[1]s_1}_{\alpha_0\alpha_1} A^{[2]s_2}_{\alpha_1\alpha_2} \cdots A^{[N]s_N}_{\alpha_{N-1}\alpha_N} |s_1 s_2 \cdots s_N\rangle, \tag{S1}$$

where $s_n \in \{0, 1\}$, $\alpha_0$ and $\alpha_N$ are one-dimensional boundary indices, and $\dim(\alpha_n) \leq \chi$. Throughout the manuscript, state fidelity means the squared overlap

$$F(|\psi_{\mathrm{MPS}}\rangle, |\psi_{\mathrm{circuit}}\rangle) = |\langle\psi_{\mathrm{MPS}}|\psi_{\mathrm{circuit}}\rangle|^2. \tag{S2}$$

We say that an MPS is in a left-, right-, or mixed-canonical form; the individual tensors in that form satisfy left- or right-isometry conditions. A right-canonical tensor satisfies

$$\sum_{s_n,\alpha_n} A^{[n]s_n}_{\alpha_{n-1}\alpha_n} A^{[n]s_n *}_{\alpha'_{n-1}\alpha_n} = \delta_{\alpha_{n-1}\alpha'_{n-1}}, \tag{S3}$$

and a left-canonical tensor satisfies

$$\sum_{\alpha_{n-1},s_n} B^{[n]s_n}_{\alpha_{n-1}\alpha_n} B^{[n]s_n *}_{\alpha_{n-1}\alpha'_n} = \delta_{\alpha_n\alpha'_n}. \tag{S4}$$

For $\chi = 2$ and local dimension $d = 2$, these tensors define isometries from a two-dimensional virtual space into a four-dimensional two-qubit space. Each such isometry can be completed to a two-qubit unitary by adding orthonormal columns.

The ladder construction uses a right-canonical MPS and prepares the state from left to right. For each site $n < N$, choose a two-qubit unitary $U^{[n]}_{\mathrm{Lad}}$ on qubits $(n, n+1)$ such that

$$\langle s_n, \alpha_n | U^{[n]}_{\mathrm{Lad}} | \alpha_{n-1}, 0\rangle = A^{[n]s_n}_{\alpha_{n-1}\alpha_n}. \tag{S5}$$

The first qubit of the block becomes the physical output $s_n$, and the second carries the propagated virtual index $\alpha_n$. The final boundary tensor is a single-qubit unitary and can be absorbed into the last two-qubit block. Thus $N-1$ nearest-neighbor two-qubit blocks exactly prepare any normalized $\chi = 2$ target MPS, and

$$F(|\psi_{\mathrm{MPS}}\rangle, |\psi_{\mathrm{circuit}}\rangle) = 1. \tag{S6}$$

The mixed-canonical construction uses the same isometry-to-unitary completion but schedules gates from the orthogonality center outward [5]. After moving the center

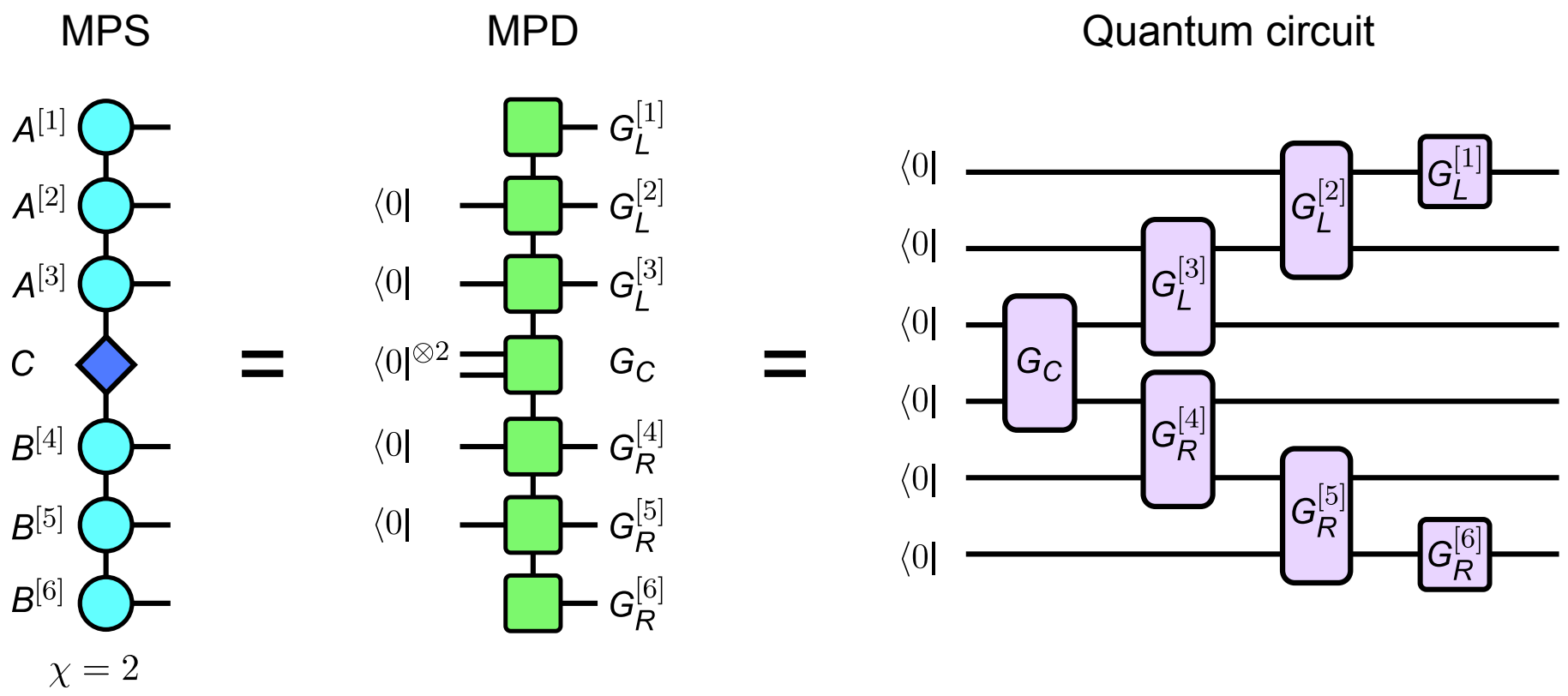


**Fig. S1** Schematic relation among a mixed-canonical $\chi = 2$ MPS, a matrix-product disentangler (MPD), and the corresponding preparation circuit. The left- and right-canonical tensors satisfy the isometry conditions of Eqs. (S4) and (S3), while the center tensor fixes the orthogonality center. The MPD disentangles the target state to $|0\rangle^{\otimes N}$; the preparation circuit uses the adjoint gate sequence. The mixed-canonical schedule is center-outward, whereas the ladder construction uses an analogous left-to-right sequence. We note that the single-qubit gates $G_L^{[1]}$ and $G_R^{[6]}$ can be absorbed into $G_L^{[2]}$ and $G_R^{[5]}$, respectively, in the circuit configuration.

to a central bond, the two-site center tensor fixes a central two-qubit unitary $U_C$ through

$$\langle \alpha_m, \beta_m | U_C | 00 \rangle = C_{\alpha_m \beta_m}. \tag{S7}$$

Left-canonical tensors to the left of the center and right-canonical tensors to the right are then embedded into two-qubit unitaries using the same column-completion argument as in Eq. (S5). Blocks at equal distance from the center act on disjoint qubit pairs and can be scheduled in parallel. Contracting the center, left-canonical, and right-canonical tensors reconstructs the original mixed-canonical MPS; equivalently, the inverse circuit disentangles the target state to $|0\rangle^{\otimes N}$. Therefore the $\chi = 2$ mixed-canonical entry in main Table 1 is also exact in the fidelity convention of Eq. (S2).

The one-layer brick-wall ansatz does not share this constructive guarantee. It has the same $N - 1$ nearest-neighbor two-qubit blocks per layer, but its fixed odd/even causal structure is different from the sequential virtual-index propagation above. Consequently, a generic $\chi = 2$ MPS is not guaranteed to have an exact one-layer brick-wall representation. The brick-wall entries in main Table 1 should therefore be read as validated iterative approximations rather than constructive exact encodings.

## 1.B Iterative construction for higher bond dimensions

For $\chi > 2$, the target MPS is approximated by $K$ architecture-specific layers of generic two-qubit unitary blocks. We write the resulting circuit state as

$$|\psi_K(\boldsymbol{\theta})\rangle = U_K(\boldsymbol{\theta}_K) \cdots U_2(\boldsymbol{\theta}_2) U_1(\boldsymbol{\theta}_1) |0\rangle^{\otimes N}, \tag{S8}$$

where $U_j$ denotes the $j$th ladder, mixed-canonical, or brick-wall architecture layer. The local sweep optimization maximizes the absolute overlap

$$O(\boldsymbol{\theta}) = |\langle \psi_{\mathrm{MPS}} | \psi_K(\boldsymbol{\theta}) \rangle|, \tag{S9}$$

and the reported main Table 1 metric is

$$F(\boldsymbol{\theta}) = |\langle \psi_{\mathrm{MPS}} | \psi_K(\boldsymbol{\theta}) \rangle|^2 = O(\boldsymbol{\theta})^2. \tag{S10}$$

For normalized states, maximizing $O$ and maximizing $F$ have the same global maximizers because the square is monotone on $[0, 1]$, although the numerical histories of the two objective functions need not be identical.

The higher-$\chi$ protocol follows the matrix-product-disentangler viewpoint used in recent MPS preparation and residual-circuit constructions [6, 7]. The local sweeps described below maximize the absolute overlap in Eq. (S9) for all three layouts. Overlap maximization is also used in Ref. [8], whereas Ref. [9] optimizes disentangling circuits through an entanglement-based objective. For $K = 1$, an architecture-specific initial circuit is formed from the target MPS: ladder layers follow the sequential right-canonical order, mixed-canonical layers use the center-outward order, and brick-wall layers use one odd and one even sublayer. The local two-qubit blocks are then optimized by sweeps. To reach $K = 2$ and $K = 3$, the previously optimized layers are retained, a new residual layer of the same architecture is appended, and the full set of layers is jointly reoptimized. This inheritance distinguishes the decomposition layer number $K$ from the later VQE layer number $L$.

Local sweeps were continued until the target-overlap tolerance was reached or the prescribed maximum sweep count was exhausted. The tolerance ranged from $10^{-5}$ to $10^{-8}$ across the decomposition cases. The final fidelities are reported in main Table 1. The exact ladder and mixed-canonical $\chi = 2$ entries follow from the constructive mappings above, while the brick-wall $\chi = 2$ and all $\chi = 8$ entries are numerical iterative approximations.

For the system-size and layer-count scans of main Fig. 4 ($\chi = 8$, $N = 10$–$50$, $K = 1$–$5$), the circuit initialized by the decomposition is refined with L-BFGS-B at each $K$ by maximizing the state fidelity $F$ using the Qiskit AQC-Tensor addon [10] and its quimb backend [11] (at most 5000 iterations; stopping when the decrease in $1 - F$ between successive iterations is below $10^{-8}$, with no gradient-norm stopping criterion). Every point is computed six times. One run starts from the unperturbed decomposition parameters; the other five add independent Gaussian noise of standard deviation 0.05 rad to the initial angles before the refinement, and the resulting layers are carried forward to the next value of $K$. Fig. 4 reports the mean over the six runs and their min–max range as error bars. The spread is small for the mixed-canonical layout and largest for the ladder layout at $N \geq 20$ and $K \geq 3$, where it reaches 0.05 in $1 - F$. Differences between neighboring system sizes that are smaller than this run-to-run spread are not resolved.

The target MPSs used for the encoding-fidelity diagnostics in main Table 1 and Fig. 4 are bond-dimension-constrained DMRG approximations for the same twisted-bond Heisenberg Hamiltonian. The target MPSs were generated in Julia using

ITensors/ITensorMPS [12], qubit site indices, a random-MPS initial state with link dimension 2, requested maximum bond dimension $\chi$, cutoff $10^{-8}$, and DMRG noise $10^{-5}$. They implement the Hamiltonian with spin operators $\mathbf{S} = \boldsymbol{\sigma}/2$, so the target tensors correspond to the same eigenstates as the Pauli-normalized Hamiltonian used in VQE and the variational energies of the Pauli-normalized Hamiltonian are four times those of the spin-operator Hamiltonian. The $N = 8$ targets of main Table 1 are converged: independent DMRG runs agree to within $5 \times 10^{-7}$ in relative energy within 30 sweeps. The $\chi = 8$ targets of Fig. 4 are not fully converged for $N \geq 30$: independent runs with 200 sweeps differ by up to $3 \times 10^{-3}$ in relative energy, and the energy is still changing after 200 sweeps. Fig. 4 therefore characterizes the encoding of one fixed bond-dimension-constrained DMRG target per system size rather than of a converged $\chi = 8$ target. The relative-energy reference $E_g$ in the VQE figures is obtained independently from exact diagonalization of the finite Pauli-normalized Hamiltonian.

The center-outward residual-layer construction used here differs from the multilayer sloping-brickwork circuits considered in Ref. [13]. In that work, sloping brickwork is formed by stacking staircase-style layers, whereas each present mixed-canonical pretrained layer is generated from a residual MPS with a chosen orthogonality center and inherited layers are jointly reoptimized. We therefore do not identify the two constructions.

## 1.C Relation to VQE initialization

The $K$ circuit layers optimized during MPS-to-circuit encoding are used as the pretrained layers of the total $L$-layer VQE circuit, while the remaining $L - K$ layers are appended according to the same architecture and initialized as specified in the main text. During VQE all circuit parameters, including the pretrained parameters, remain trainable. The same interpretation of $K$ and $L$ is used for the ideal, noisy, ZNE-mitigated, and distributed-QPU results. For the distributed circuits at $L = 3$ in Fig. 7, $N = 9$ qubits are partitioned over $M = 3$ QPUs as $\{0, 1, 2\}$, $\{3, 4, 5\}$, and $\{6, 7, 8\}$, with communication qubits 0, 3, and 6. The NN distributed ansatz uses inter-QPU edges $(0, 3)$ and $(3, 6)$, while the all-to-all ansatz also includes $(0, 6)$. The target MPS has $\chi = 8$. Two distributed architecture layers ($K = 2$) are initialized near identity and optimized jointly, including both local and inter-QPU blocks, against the target MPS using the iterative MPS-to-circuit procedure. These optimized layers initialize VQE; one additional near-identity-initialized architecture layer (Methods) is appended to obtain a total layer number $L = 3$, and all pretrained and appended parameters remain trainable. This distributed pretraining is an iterative approximation protocol, not a constructive exact encoding. For the distributed simulations (Figs. 7 and 8 and the reference calculations of Sec. 3.H) the local sweeps use untruncated environments (environment bond dimension 16, exact for $N = 9$) and at most 500 sweeps; each of the five runs of an ensemble is initialized by an independent MPS pretraining run, and the noiseless, noisy, and ZNE-assisted runs of a circuit, at every noise setting, start from the same five parameter sets (Sec. 3.B). The encoding fidelities reached by the $K = 2$ pretraining stage in these ensembles (five independent pretraining runs per circuit; individual encoding fidelities and random-number seeds are included in the source data) are 0.767(17) for the brick-wall chain, 0.84(4) for the distributed NN

layout, 0.966(5) for the all-to-all layout, and 0.9582(5) for the two-QPU link circuit (parentheses: sample standard deviation in the last digits); the NN pretraining runs fall into two groups (0.80 and 0.86–0.89); in unmitigated noisy VQE at $t_{2q} = 50\,\mu$s and $r = 1$ [$L = 3$ in Fig. 7(c)], the final relative energy errors are 0.23 for the two runs of the lower group and 0.17–0.19 for the other three.

## 2 Circuit resources and distributed scheduling

The resource quantities in Fig. 2, Fig. 7, and main Table 1 count abstract two-qubit unitary blocks rather than hardware-native pulse time. Let $G_2$ denote the number of two-qubit blocks per architecture layer and $D_2$ the two-qubit-block depth, where mutually disjoint two-qubit blocks in the same round contribute one unit of depth. Each generic two-qubit block contains 15 trainable parameters, so the parameter count is $15G_2$ [14, 15]. Local single-qubit rotations are included in this block parameterization and are not counted as separate blocks in Fig. 2.

For the main ansatz families, every architecture layer contains the same nearest-neighbor block set, so

$$G_2^{\mathrm{Lad}} = G_2^{\mathrm{MC}} = G_2^{\mathrm{BW}} = N - 1. \tag{S11}$$

The depths differ because the schedules differ. The ladder construction propagates the virtual index sequentially, giving $D_2^{\mathrm{Lad}} = N - 1$. The mixed-canonical construction proceeds from the central bond outward, allowing left and right blocks at the same distance from the center to run in parallel, so $D_2^{\mathrm{MC}} = \lceil N/2 \rceil$. A brick-wall layer consists of one boundary-starting sublayer and one shifted sublayer, so $D_2^{\mathrm{BW}} = 2$.

For the distributed-QPU construction, the $N$ qubits are divided across $M$ QPUs with one designated communication qubit per QPU. Local intra-QPU brick-wall blocks have depth 2 and run in parallel across QPUs. Inter-QPU communication blocks are then scheduled as matchings on the QPU coupling graph, because a QPU with one communication qubit cannot participate in two simultaneous inter-QPU blocks. For a nearest-neighbor QPU path, the $M - 1$ inter-QPU blocks require

$$R_{\mathrm{NN}}(M) = \begin{cases} 1, & M = 2, \\ 2, & M \geq 3, \end{cases} \qquad D_{2,\mathrm{NN}} = 2 + R_{\mathrm{NN}}. \tag{S12}$$

For all-to-all QPU coupling, the complete graph $K_M$ has $M(M-1)/2$ inter-QPU blocks. Edge coloring under the same single-qubit matching constraint gives

$$R_{\mathrm{all}}(M) = \begin{cases} M - 1, & M \text{ even}, \\ M, & M \text{ odd}, \end{cases} \qquad D_{2,\mathrm{all}} = 2 + R_{\mathrm{all}}. \tag{S13}$$

Thus the nearest-neighbor distributed depth is 3 for $M = 2$ and 4 for $M \geq 3$, whereas the all-to-all depth is $M + 1$ for even $M$ and $M + 2$ for odd $M$.

**Table S1** Main and distributed circuit resources per architecture layer. $D_2$ is abstract two-qubit-block depth; it is not a physical runtime and does not assume equal native durations for all hardware operations.

| Construction | Two-qubit blocks | Block depth or communication rounds |
|---|---|---|
| Ladder | $G_2 = N - 1$ | $D_2 = N - 1$ |
| Mixed-canonical | $G_2 = N - 1$ | $D_2 = \lceil N/2 \rceil$ |
| Brick-wall | $G_2 = N - 1$ | $D_2 = 2$ |
| Distributed NN path | local blocks plus $M - 1$ inter-QPU blocks; total $N - 1$ | $R_{\mathrm{NN}} = 1$ for $M = 2$, 2 for $M \geq 3$; total $D_2 = 3$ or 4 |
| Distributed all-to-all | $N - M + M(M - 1)/2$ | $R_{\mathrm{all}} = M - 1$ for even $M$, $M$ for odd $M$; total $D_2 = M + 1$ or $M + 2$ |

The phrase "constant per-layer block depth" in the main text refers to the scaling of $D_2$, the two-qubit-block depth of one isolated architecture layer, under these block-scheduling conventions. For a single brick-wall layer, $D_2 = 2$ is independent of the chain length $N$. For the distributed nearest-neighbor schedule, $D_2 = 4$ for all $M \geq 3$ under the one-communication-qubit path model. In contrast, all-to-all communication depth grows approximately linearly with $M$.

The depth of a multilayer circuit is determined by its complete gate schedule and need not equal the number of layers multiplied by $D_2$. For the unfolded circuits in main Figs. 5 and 6 with $N = 8$ and $(L, K, \chi) = (4, 3, 8)$, the ladder, mixed-canonical, and brick-wall implementations have two-qubit-block depths of 13, 10, and 8, respectively (11, 8, and 6 at $(3, 2, 8)$): the ladder and mixed-canonical schedules overlap operations from successive layers while preserving their dependencies. For the ladder and brick-wall schedules the unfolded depths are $N + 2L - 3$ and $2L$, so that the ladder exceeds the brick-wall by $N - 3$ blocks at any $L$. The brick-wall implementations, including the distributed circuits in main Figs. 7 and 8, use layer-by-layer schedules, for which the unfolded VQE block depth is $LD_2$. The encoding fidelities in main Table 1 and Fig. 4 are evaluated classically from overlaps of ideal circuit states with the target MPS and do not prescribe an execution schedule. Related results on logarithmic-depth approximate preparation of normal MPS [16] and constant-depth adaptive preparation of suitable MPS families using measurements and feedforward [17] concern the total state-preparation depth under their respective assumptions.

In the noisy distributed simulations, intra-QPU and inter-QPU two-qubit operations use the same gate durations and noise parameters as a baseline assumption; real modular hardware may assign different durations or error rates. The NN-DQC and all-to-all-DQC circuits in Fig. 7 are not gate- or parameter-matched: for $M = 3$, the all-to-all layer includes an additional inter-QPU block, additional trainable parameters, and greater communication depth. The comparison is therefore a capacity–resource trade-off rather than an isolation of connectivity alone. The equal block counts of the brick-wall and distributed-NN layers moreover do not imply equal coupling graphs: partitioning replaces the chain bonds $(2, 3)$ and $(5, 6)$ by the communication links $(0, 3)$ and $(3, 6)$. For the $N = 9$ twisted Hamiltonian, the summed graph distance of the eight interaction pairs is 36 on the chain, 26 on the NN graph, and 21 on the all-to-all graph, and $(3, 6)$ is itself an interaction pair of the Hamiltonian. The two-QPU circuit of Fig. 8 is the $M = 2$ case of the NN path ($D_2 = 3$) with the communication qubits chosen as 3 and 6, so that its single link $(3, 6)$ replaces the chain bond $(3, 4)$; the chain-bond control uses $(3, 4)$ instead and therefore has the coupling graph of the brick-wall at $D_2 = 3$ (Sec. 3.H).

Which interaction graphs a chain of nearest-neighbor blocks can realize follows from a coverage count. An edge of the interaction graph is covered when its two qubits are adjacent in the qubit order, so a single chain covers at most $N$ minus the minimum number of vertex-disjoint paths needed to cover the graph; a path graph is covered completely by a suitable order, whereas a tree with one branch point of degree three loses one edge, a simple cycle loses one edge, and the $3 \times 3$ lattice loses four. An exhaustive enumeration over all $9!/2$ qubit orders for $N = 9$ (and, for the modular layouts, over all choices of one communication qubit per QPU) gives: the open chain

and the twisted model, which is a relabeled chain, are covered 8/8 by a single chain, 8/8 by the two-QPU layout with one link, and 7/8 by the three-QPU NN and all-to-all layouts, which lose one chain bond per cut that the single communication qubit cannot restore; a tree with one branch point (arms of 3, 3, and 2 qubits) is covered 7/8 by a chain but 8/8 by the three-QPU layouts, which place the communication qubit on the branch point; and the $3 \times 3$ square lattice (12 edges) is covered 8/12 by a snake-ordered chain and also 8/12 by the two- and three-QPU layouts, so no connectivity gain is available there. For the twisted model in the fixed MPS order of the main text, the single chain covers only the bond $(4, 5)$, and the inter-QPU links of the distributed circuits realize further interaction pairs. Whether the additional coverage at a branch point lowers the noisy energies is tested in Sec. 3.J.

# 3 Numerical protocols and noise model

## 3.A Simulation configurations and statistical summaries

The VQE calculations use the Adam optimizer with an initial learning rate of $10^{-2}$ and the learning-rate reduction and gradient clipping described in Methods, except for the illustrative noiseless trajectory in Fig. 1(b), which uses L-BFGS. The $N = 8$ noiseless VQE trajectories are run for $2 \times 10^4$ iterations, whereas noisy and ZNE-mitigated trajectories, and all $N = 9$ trajectories, are run for $10^4$ iterations. All VQE optimizations use exact expectation values from state-vector or density-matrix simulation; measurement shot noise is not included. Tables S2 and S3 collect the run counts and displayed statistics. Bands and error bars describe run-to-run or sample-to-sample variability; they are not standard errors unless explicitly stated.

## 3.B Pretraining protocols

The single-processor and distributed calculations use different pretraining protocols, and their ensembles are therefore not pooled. For the $N = 8$ single-processor VQE results (Figs. 3(a)–(c), 5, and 6, and Fig. S2) the pretraining is run once per circuit configuration, and the runs of an ensemble differ only in the random initialization of the appended layers; the noiseless, noisy, and ZNE-assisted ensembles come from separate pretrainings. At $(L, K, \chi) = (4, 3, 8)$, the noisy and ZNE-assisted calculations follow the same protocol as at $(3, 2, 8)$. The noiseless references are those of Fig. 3. The gradient diagnostics of Fig. 3(d),(e) use Haar-random blocks without pretraining, and the encoding scans of Fig. 4 follow the protocol of Sec. 1. For the $N = 9$ distributed calculations (Figs. 7 and 8, Table 2, Tables S9–S13, and Secs. 3.I and 3.J) the local sweeps use untruncated environments (bond dimension 16, exact for $N = 9$) and at most 500 sweeps. The five runs of an ensemble are initialized by five independent MPS pretraining runs using random-number seeds $s_0, \ldots, s_0{+}4$ ($s_0 = 130000$ for Figs. 7 and 8); the resulting parameter sets are reused by the noiseless, noisy, and ZNE-assisted runs of the same circuit at every noise setting, so that the training methods are paired by their initial parameters. Different circuits do not share initial states. Comparisons between circuits use their means and sample standard deviations, as described in the main-text Methods. The link scan of Table S11 uses the distinct random-number seeds

**Table S2** Numerical configurations and statistical summaries for the ideal-circuit and encoding results of the main text (Figs. 1(b), 3, and 4). SD denotes the sample standard deviation with Bessel's correction, and min–max denotes the full range across runs.

| Panel | Configuration | Evaluation budget | Reported quantity |
|---|---|---|---|
| *Overview diagnostic* | | | |
| Fig. 1(b) | $N = 8$; random-circuit layer count $1, \dots, 10$ | 10,000 circuit samples per layer | Mean purity; sample-SD bars |
| *Ideal-circuit and encoding diagnostics* | | | |
| Fig. 3(a)–(c) | Ideal VQE with $N = 8$, $L = 4$, and $(K, \chi) = (1, 2), (3, 4), (3, 8)$ | 20 runs per ansatz and setting, from one MPS pretraining run; 20,000 iterations per run | Mean relative energy error; min–max band |
| Fig. 3(d)–(e) | (d) varying $N$ at $L = 3$; (e) varying $L$ at $N = 10$ | 100 circuits per point with Haar-random SU(4) blocks (no pretraining) | Unbiased sample variance of the selected `RYY` gradient |
| Fig. 4(a) | $\chi = 8$, $K = 4$; $N = 10, 20, 30, 40, 50$ | Six restarts per ansatz and $N$ | Mean state infidelity $1 - F$; error bars = min–max |
| Fig. 4(b) | $\chi = 8$, $N = 50$; $K = 1, \dots, 5$ | Six restarts per ansatz and $K$ | Mean state infidelity $1 - F$; error bars = min–max |

140000–140004, so its reference values differ from those of Table S12. For the $N = 8$ single-processor calculations, the environment-truncation setting of three brick-wall pretraining runs, those of the ZNE-assisted ensembles of Fig. 6 and Fig. S2 and of the noiseless reference at $(L, K, \chi) = (3, 2, 8)$, is not recorded; the other $N = 8$ VQE ensembles start from pretraining with untruncated environments. A truncation of the sweep environments to bond dimension 8 changes the brick-wall encoding fidelity by more than $10^{-3}$ in one of ten tested pretraining runs at $K = 2$ and in eight of ten at $K = 3$.

*Paired-protocol control.* For the control calculations of Sec. 3.E, each single-processor circuit at $(L, K, \chi) = (3, 2, 8)$ and $(4, 3, 8)$ has five stored initial-parameter sets (random-number seeds 160000–160004), produced with the pretraining settings of the distributed calculations (untruncated environments, at most 500 sweeps per optimization stage) and reused by the noiseless, noisy, and ZNE-assisted runs, so that the three training methods are paired sample by sample as in Figs. 7 and 8. The ladder and mixed-canonical pretraining contains no random draw: the sequential decomposition of the target MPS is deterministic up to the numerical non-reproducibility of its sweeps, so the five samples of these circuits differ in the sweep results and in the seeded near-identity initialization of the appended layers, whereas the brick-wall samples are independent seeded draws. The mean pretraining fidelities are 0.813–0.815 at $(3, 2, 8)$ and 0.989–0.992 at $(4, 3, 8)$ (Table S7).

## 3.C Decomposed-gate noise model and scheduling

The noisy simulations combine a biased Pauli channel with a thermal-relaxation channel after each decomposed gate occurrence. Each generic 15-parameter SU(4) block on qubits $(a, b)$ is first decomposed into single-qubit rotations and Ising interactions. We use $R_\mu(\theta) = \exp[-i\theta\sigma_\mu/2]$ for single-qubit rotations and $\text{Ising}_{\mu\mu}(\theta) = \exp[-i\theta\sigma_a^\mu\sigma_b^\mu/2]$. The parameter vector is ordered as

$$\begin{aligned}&R_z^a(\theta_1), R_y^a(\theta_2), R_z^a(\theta_3), R_z^b(\theta_4), R_y^b(\theta_5), R_z^b(\theta_6),\\ &\text{Ising}_{XX}(\theta_7), \text{Ising}_{YY}(\theta_8), \text{Ising}_{ZZ}(\theta_9),\\ &R_z^a(\theta_{10}), R_y^a(\theta_{11}), R_z^a(\theta_{12}), R_z^b(\theta_{13}), R_y^b(\theta_{14}), R_z^b(\theta_{15}).\end{aligned} \tag{S14}$$

Adjacent single-qubit rotations generated by neighboring blocks or layers are not analytically merged before the noisy simulation. For a scheduled set of disjoint two-qubit blocks, decomposed gates occupying the same position in Eq. (S14) are grouped into the same gate sublayer when their supports are disjoint; idle wires then relax for the duration assigned to that gate sublayer. The noise model is applied to these gate operations, not once to the undecomposed block. For a single-qubit gate, the biased Pauli error probability is $p_s = 10^{-5}$. For a two-qubit gate, independent copies of the same single-qubit biased Pauli channel act on the two active qubits with probability $p_d = 10^{-4}$ per active qubit:

$$\mathbb{E}_{P,p_d}^{(2)} = \mathbb{E}_{P,p_d} \otimes \mathbb{E}_{P,p_d}. \tag{S15}$$

For each gate class, $p_X = p_Y = 0.001p$ and $p_Z = 0.998p$, where $p$ is $p_s$ or $p_d$ for the corresponding active qubit. Thus $p_d$ is not a total two-qubit-block error probability; the probability of at least one non-identity Pauli event during a two-qubit gate occurrence is $1 - (1 - p_d)^2 \simeq 1.9999 \times 10^{-4}$. The model does not add an independent 15-outcome correlated two-qubit Pauli channel, although simultaneous independent errors can produce two-qubit Pauli strings. All noisy results use the combined biased-Pauli and thermal-relaxation channel; we do not interpret the total error as an additive decomposition of the two contributions.

For a gate or idle duration $\tau$, the single-qubit thermal-relaxation component is parameterized by $T_1$ and $T_2$ [18, 19]. In the zero-temperature form used here,

$$\rho_{11} \mapsto e^{-\tau/T_1}\rho_{11}, \tag{S16}$$

$$\rho_{00} \mapsto \rho_{00} + \left(1 - e^{-\tau/T_1}\right)\rho_{11}, \tag{S17}$$

$$\rho_{01} \mapsto e^{-\tau/T_2}\rho_{01}. \tag{S18}$$

Equivalently, this can be represented by a Kraus channel with reset probability $1 - e^{-\tau/T_1}$ and dephasing chosen to reproduce the coherence decay. For an active single-qubit gate, the ideal gate is followed by the biased Pauli channel and then by thermal relaxation for the gate duration. For an active two-qubit gate, the biased Pauli channel

acts independently on the two active qubits and is followed by tensor-product single-qubit thermal relaxation. The simulation uses $t_{1\mathrm{q}} = 10\,\mathrm{ns}$ for single-qubit gates and $t_{2\mathrm{q}} = 50\,\mu\mathrm{s}$ for two-qubit gates. Idle qubits receive thermal relaxation for the duration of the current scheduled sublayer: $t_{1\mathrm{q}}$ for a single-qubit-only sublayer and $t_{2\mathrm{q}}$ for any sublayer containing a two-qubit gate. Folded inverse gates use the same gate durations. For a qubit that has not yet been acted on and therefore remains in the initial $|0\rangle$ state, the zero-temperature thermal-relaxation channel leaves the state unchanged, since there is neither excited-state population to relax nor coherence to dephase. Idle-time decoherence therefore becomes relevant once nontrivial populations or coherences have been generated by preceding gates.

*Comparison with IBM fake-backend calibration snapshots.* Table S5 compares the present model with FakeSherbrooke, FakeBrisbane, FakeTorino, and FakeFez from Qiskit IBM Runtime 0.46.1, using snapshots dated 26 February 2025 [20]. The IBM gate errors in the table are reported calibration parameters. Qiskit Aer 0.17.2 interprets these parameters as average gate infidelities and chooses a depolarizing strength after accounting for thermal relaxation, subject to physical bounds [19]. If thermal relaxation alone has a larger infidelity than the reported error, no depolarizing component is added; the generated channel therefore need not reproduce the reported error. When constructing the relaxation channel, Qiskit Aer also truncates $T_2$ to $2T_1$ where necessary.

Both constructions apply gate errors before thermal relaxation, following the ideal gate. For two-qubit operations, Qiskit Aer combines a two-qubit depolarizing channel with a tensor product of single-qubit relaxation channels, whereas the present model uses independent biased Pauli channels on the two active qubits. With `NoiseModel.from_backend`, idle relaxation is applied to explicit delay instructions, including those inserted by scheduling; here it is applied to every idle wire in each scheduled sublayer.

Each generic two-qubit block in our implementation contains three Ising rotations, each followed by its own noise channel. For comparison, exact synthesis of a general two-qubit unitary on a connected qubit pair can use at most three ECR or CZ gates, together with single-qubit gates, and three are generically necessary [14, 15]; the gate count of a transpiled circuit additionally depends on optimization and routing. The tabulated two-qubit errors refer to individual elementary gates, not to complete generic blocks. In all four fake-backend targets, $R_z$ rotations have zero duration and zero assigned error, whereas our model assigns them the same duration and noise as other single-qubit rotations.

## 3.D Single-processor comparison at $(L, K, \chi) = (3, 2, 8)$

Main-text Figs. 5(b) and 6 compare the three single-processor layouts at $(L, K, \chi) = (4, 3, 8)$, at which their noiseless VQE errors are comparable (Table S6). At $L = 3$ this is not the case: the ladder and mixed-canonical layouts reach final noiseless errors of about 0.07, whereas the brick-wall reaches 0.021(13), and the same gap appears for $(K, \chi) = (1, 2)$. Figure S2 shows noisy and ZNE-assisted VQE at $(L, K, \chi) = (3, 2, 8)$. The brick-wall again has the lowest noisy error and, after ZNE-assisted training, the lowest extrapolated and noiseless errors. Its mean noisy error is lower than that of

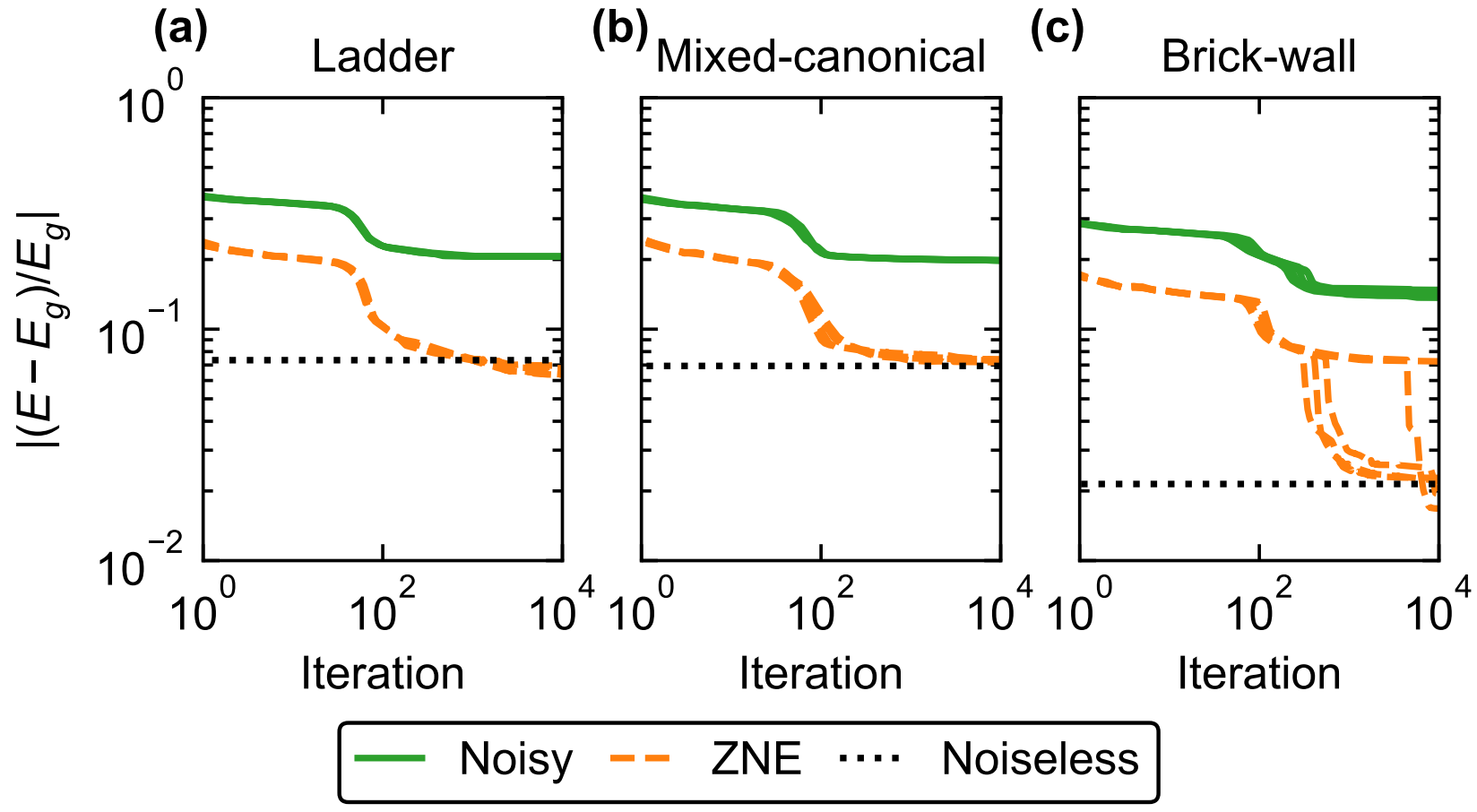


**Fig. S2** Noisy VQE and ZNE-assisted VQE at $(L, K, \chi) = (3, 2, 8)$ for the (a) ladder, (b) mixed-canonical, and (c) brick-wall ansatzes ($N = 8$, $t_{2q} = 50\,\mu$s): relative energy error $|(E - E_g)/E_g|$ of five runs each for noisy VQE (green solid) and ZNE-assisted VQE (orange dashed, extrapolated energy). The black dotted line indicates the mean final relative energy error over 20 noiseless VQE runs of the same architecture at the same setting (20,000 Adam iterations).

the ladder by 0.0653: 0.0497 comes from the lower noiseless error at the noisy-trained parameters and 0.0156 from the smaller noise-induced shift. At $(4, 3, 8)$ the brick-wall instead ends noisy training with a higher mean noiseless error than the other layouts (two runs at 0.015 and three at 0.075), and its lower noisy error comes from a smaller shift. Evaluated under the same noise without further training, the noiseless-trained parameters of Fig. 3 at $(4, 3, 8)$ give shifts of 0.256(13), 0.221(14), and 0.232(12) for the ladder, mixed-canonical, and brick-wall layouts, whereas the noisy-trained ensembles give 0.175, 0.174, and 0.108. After noisy training, the brick-wall shift is smaller than those of the other layouts, consistent with an adaptation to noise. The noiseless and noisy ensembles, however, start from separate pretrainings. In both settings the ZNE-assisted ladder runs end with a negative extrapolation error $b$ in all five runs; at $(4, 3, 8)$ three of them end below $E_g$ (Sec. 4).

## 3.E Paired-protocol and gate-error-dominated controls

Table S7 repeats the single-processor comparison of Sec. 3.D and main-text Figs. 5–6 with the paired protocol of Sec. 3.B. The ordering of the three layouts is that of Table S6: the brick-wall has the lowest noisy error and, after ZNE-assisted training, the lowest noiseless error at both settings, and the extrapolated energy of the ladder ends below $E_g$ in four of five runs at $(4, 3, 8)$. The brick-wall again has the smallest noise-induced shift, 0.1424(27) versus 0.173(3) and 0.1729(20) at $(4, 3, 8)$; the difference is smaller than in Table S6 because the noisy-trained brick-wall circuits reach a lower noiseless error here, 0.008(4) versus 0.051(33), with a correspondingly larger shift. At $(3, 2, 8)$ the ZNE-assisted brick-wall runs again end in two groups, with noiseless errors near 0.02 and 0.07.

*Gate-error-dominated control.* To test whether the smaller noise-induced shift of the brick-wall reflects its shorter schedule rather than its gate count, the three $(3, 2, 8)$ circuits were optimized under noise with the two-qubit biased-Pauli probability raised to $p_d = 5 \times 10^{-3}$ per active qubit at $t_{2\mathrm{q}} = 5\,\mu$s, all other settings and the initial parameters being those of the paired protocol. The average two-qubit-gate infidelity is then $8.0 \times 10^{-3}$ from the Pauli channel alone and $4.0 \times 10^{-4}$ from the relaxation alone, so the gate error dominates by a factor of about twenty, and the three circuits contain the same number of two-qubit gates (63). Table S8 compares the noise-induced shifts with those of the paired runs at $50\,\mu$s. Under gate-error-dominated noise the brick-wall no longer has the smallest shift, whereas at the baseline duration its shift is the smallest. The brick-wall keeps the lowest noisy error in the control because its noiseless error at $(3, 2, 8)$ is lower, not because of a smaller shift; equal gate counts do not make the Pauli contribution layout-independent, since it depends on the gate order and on the state at each gate.

## 3.F Sensitivity to the two-qubit-gate duration

To test how the baseline two-qubit-gate duration affects unmitigated noisy VQE, we repeated the single-processor noisy calculation at $(L, K, \chi) = (3, 2, 8)$ (Sec. 3.D) while varying only $t_{2\mathrm{q}}$. The tested values were $5\,\mu$s, $50\,\mu$s, and $500\,\mu$s. All other Hamiltonian, ansatz, pretraining, optimizer, biased-Pauli, thermal-relaxation, and idle-scheduling settings were fixed. Five runs with distinct random-number seeds were performed for each architecture and duration, each for 10,000 iterations. The primary metric is the final absolute relative energy error; as a convergence robustness check we also computed the mean absolute relative energy error over the final 1000 iterations. The $50\,\mu$s calculation is independent of the $(3, 2, 8)$ runs of Sec. 3.D and uses the same noise parameters.

Figure S3 shows the raw optimization trajectories at the baseline $50\,\mu$s duration together with the final errors across the tested durations. The final error and its mean over the last 1000 iterations give the same architecture ordering at each tested duration. The brick-wall ansatz has the lowest mean final error at all three durations, while the ladder/mixed-canonical ordering changes with $t_{2\mathrm{q}}$ and the absolute error scale increases as $t_{2\mathrm{q}}$ increases. These results show that the favorable brick-wall ordering persists across the tested variation in $t_{2\mathrm{q}}$.

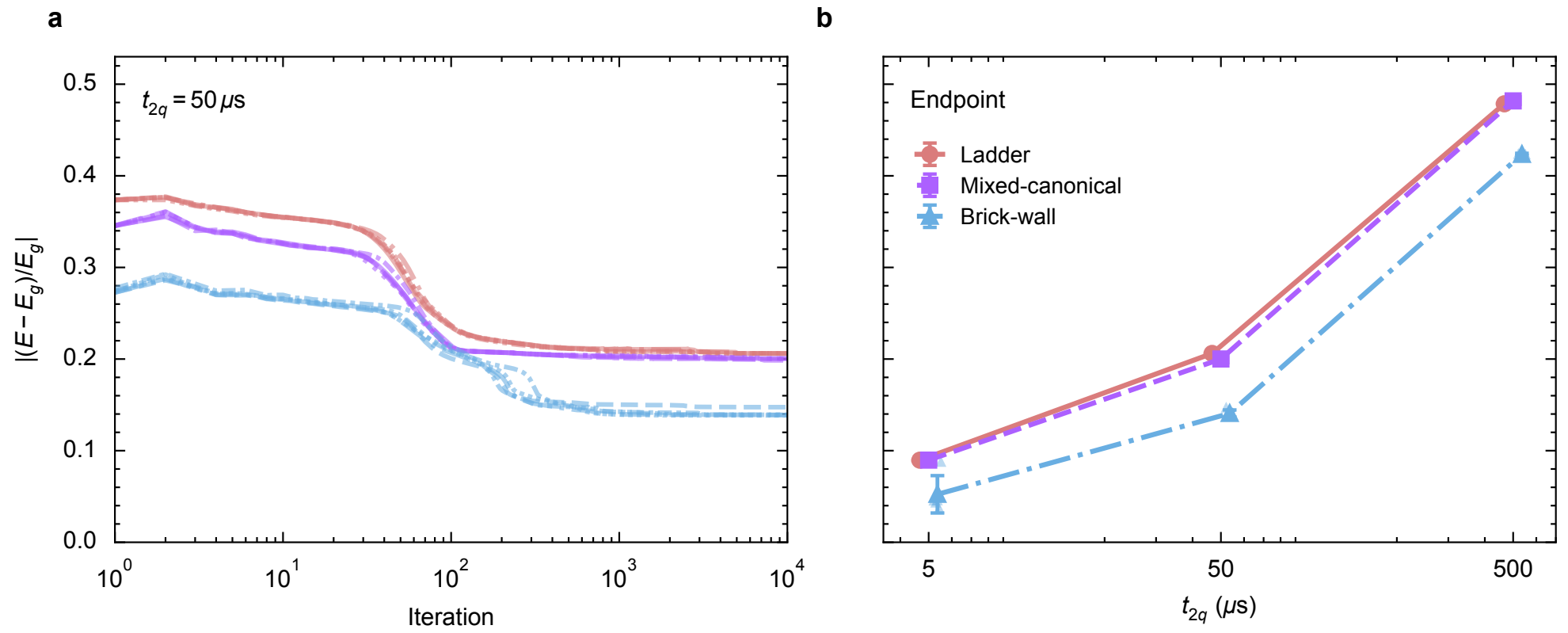


**Fig. S3** Duration sensitivity of unmitigated noisy VQE. (a) Raw optimization trajectories of the relative energy error $|(E - E_g)/E_g|$ for five runs per architecture at the baseline $t_{2\mathrm{q}} = 50\,\mu$s. (b) Final relative energy error for five runs at each two-qubit-gate duration. Points show individual runs, while markers and error bars denote the mean and sample standard deviation. All runs use the single-processor noisy-VQE setting with $N = 8$ and $(L, K, \chi) = (3, 2, 8)$; only $t_{2\mathrm{q}}$ is varied. The brick-wall ansatz has the lowest mean final error at all three tested durations, while the ladder/mixed-canonical ordering changes with duration and the error scale increases with $t_{2\mathrm{q}}$.

**Table S3** Numerical configurations and statistical summaries for the noisy and ZNE-assisted VQE results (Figs. 5–8 and Table 2 of the main text, Fig. S2, and Tables S7, S8, and S16). SD denotes the sample standard deviation with Bessel's correction.

| Panel | Configuration | Evaluation budget | Reported quantity |
|---|---|---|---|
| Figs. 5(b) and 6 | $N = 8$; $(L, K, \chi) = (4, 3, 8)$; noiseless VQE references (runs of Fig. 3) | 20 runs per ansatz; 20,000 iterations | Mean final relative energy error: ladder in Fig. 5(b), each architecture in Fig. 6 |
| Figs. 5(b) and 6 | $N = 8$; $(L, K, \chi) = (4, 3, 8)$; noisy and ZNE-assisted VQE | 5 trajectories per ansatz and method, from one MPS pretraining run; 10,000 iterations | Individual noisy and, where shown, ZNE trajectories (extrapolated and noiseless energies) |
| Fig. S2 | $N = 8$; $(L, K, \chi) = (3, 2, 8)$; noiseless references, noisy and ZNE-assisted VQE | Noiseless: 20 runs, 20,000 iterations; noisy and ZNE-assisted: 5 runs, 10,000 iterations | Individual noisy and ZNE trajectories; mean final noiseless error |
| Fig. 7(c)–(f) | $N = 9$; panel (c) varies $L = 1, \ldots, 4$ with $(L, K, \chi) = (1, 1, 2)$, $(2, 1, 8)$, $(3, 2, 8)$, $(4, 3, 8)$; panels (d)–(f) use $(L, K, \chi) = (3, 2, 8)$ | 5 trajectories per plotted family and method, each initialized by an independent MPS pretraining run; in panels (d)–(f) the resulting initial parameters are shared by all methods (Sec. 3.B); 10,000 iterations | (c) Mean final relative energy error with sample-SD bars; (d)–(f) individual trajectories |
| Fig. 8 | $N = 9$; $(L, K, \chi) = (3, 2, 8)$; $r = r_t = r_p \in \{1, 2, 5, 10\}$; $t_{2q} = 50$ and $5\,\mu$s | 5 trajectories per point and method, each initialized by an independent MPS pretraining run, with the resulting initial parameters shared by all methods and noise settings; 10,000 iterations | Mean final relative energy error with sample-SD bars; individual brick-wall runs |
| Table 2 | $N = 9$; $(L, K, \chi) = (3, 2, 8)$; pretrained start and noiseless, noisy, and ZNE-assisted VQE of the circuits of Fig. 8, the VQE runs evaluated with noiseless, noisy, and extrapolated energies, respectively; noiseless energies at the trained parameters as diagnostics | 5 runs per circuit and training method | Mean final relative energy error with sample SD |
| Tables S7 and S8 | $N = 8$; $(L, K, \chi) = (3, 2, 8)$ and $(4, 3, 8)$; noiseless, noisy, and ZNE-assisted VQE from five stored initial-parameter sets per circuit; GE: the $(3, 2, 8)$ circuits at $t_{2q} = 5\,\mu$s with $p_d = 5 \times 10^{-3}$ | 5 runs per circuit and method; 20,000 (noiseless) or 10,000 iterations | Mean final relative energy error with sample SD; noise-induced shift |
| Table S16 | $N = 9$; $(L, K, \chi) = (3, 2, 8)$; $r = 1$; ZNE estimate at the final parameters of the noisy-VQE runs of Fig. 8 | one evaluation per run (fold factors 1, 3, 5) | Mean relative energy error with sample SD |

**Table S4** Compact noise and ZNE settings for the noisy main-text simulations. The fold factor $\lambda$ is a circuit-fold label, not an exact scalar multiplier of the physical noise channel.

| Quantity | Value | Applies to | Note |
|---|---|---|---|
| Single-qubit Pauli probability | $p_s = 10^{-5}$ | Figs. 5–7 noisy and ZNE simulations | $p_X = p_Y = 0.001p_s$, $p_Z = 0.998p_s$. |
| Two-qubit Pauli probability | $p_d = 10^{-4}$ per active qubit | Figs. 5–7 noisy and ZNE simulations | independent channel copies on the two active qubits; total non-identity probability $1 - (1 - p_d)^2$. |
| Gate durations | $t_{1\mathrm{q}} = 10\,\mathrm{ns}$, $t_{2\mathrm{q}} = 50\,\mu\mathrm{s}$ | active gates, folded inverse gates, and idle intervals | idle wires use the current scheduled sublayer duration; Fig. 8(b),(d) use $t_{2\mathrm{q}} = 5\,\mu\mathrm{s}$. |
| Thermal relaxation | $T_1 = 500\,\mathrm{ms}$, $T_2 = 10\,\mathrm{ms}$ | active gates and idle intervals | Intra-QPU and inter-QPU two-qubit operations use the same durations and noise parameters in Fig. 7; Fig. 8 multiplies the inter-QPU duration and Pauli probability by $r$. |
| ZNE folds | $\lambda = 1, 3, 5$ | Figs. 6, 7(d)–(f), and 8(c),(d) | global unitary folding of the complete parameterized circuit. |
| ZNE extrapolation | quadratic estimator at $\lambda = 0$ | Figs. 6, 7(d)–(f), and 8(c),(d) | weights 15/8, −5/4, and 3/8 for folds 1, 3, and 5. |

**Table S5** Noise parameters of the present model and calibration snapshots of four IBM fake backends. For IBM, $T_1$, $T_2$, single-qubit-gate parameters, and readout errors are medians over qubits; two-qubit-gate parameters are medians over supported qubit pairs, counting each pair once. The IBM single-qubit entries refer to SX gates; X gates have the same tabulated medians, while virtual $R_z$ gates are excluded. For the present model, gate errors are $1 - F_{\text{avg}}$ of the combined Pauli and thermal-relaxation channel; IBM gate errors are the reported calibration values. The ratio $t_{2q}/T_2$ and the idle infidelity are calculated from the unrounded medians of the reported $T_1$, $T_2$, and two-qubit-gate duration, before any truncation of individual $T_2$ values; they are estimates at median parameters rather than medians of the individual ratios or infidelities. Readout error is the mean of the two conditional assignment-error probabilities, median-aggregated over qubits.

| | This work | | Sherbrooke | Brisbane | Torino | Fez |
|---|---|---|---|---|---|---|
| | 50 $\mu$s | 5 $\mu$s | (Eagle) | (Eagle) | (Heron) | (Heron) |
| $T_1$ | 500 ms | | 278 $\mu$s | 232 $\mu$s | 185 $\mu$s | 145 $\mu$s |
| $T_2$ | 10 ms | | 170 $\mu$s | 150 $\mu$s | 141 $\mu$s | 88 $\mu$s |
| 1q gate duration | 10 ns | | 57 ns | 60 ns | 32 ns | 24 ns |
| 1q gate error | $7.0 \times 10^{-6}$ | | $2.4 \times 10^{-4}$ | $2.4 \times 10^{-4}$ | $2.7 \times 10^{-4}$ | $2.3 \times 10^{-4}$ |
| 2q gate | Ising rotation | | ECR | ECR | CZ | CZ |
| 2q gate duration | 50 $\mu$s | 5 $\mu$s | 533 ns | 660 ns | 68 ns | 84 ns |
| 2q gate error | $4.2 \times 10^{-3}$ | $5.6 \times 10^{-4}$ | $7.8 \times 10^{-3}$ | $7.7 \times 10^{-3}$ | $4.2 \times 10^{-3}$ | $3.9 \times 10^{-3}$ |
| $t_{2q}/T_2$ | $5.0 \times 10^{-3}$ | $5.0 \times 10^{-4}$ | $3.1 \times 10^{-3}$ | $4.4 \times 10^{-3}$ | $4.8 \times 10^{-4}$ | $9.6 \times 10^{-4}$ |
| Idle infidelity per 2q gate | $1.7 \times 10^{-3}$ | $1.7 \times 10^{-4}$ | $1.4 \times 10^{-3}$ | $1.9 \times 10^{-3}$ | $2.2 \times 10^{-4}$ | $4.1 \times 10^{-4}$ |
| Readout error | none | | $2.0 \times 10^{-2}$ | $2.0 \times 10^{-2}$ | $2.3 \times 10^{-2}$ | $7.6 \times 10^{-3}$ |

**Table S6** Single-processor layouts at $(L, K, \chi) = (3, 2, 8)$ and $(4, 3, 8)$, $N = 8$, $t_{2q} = 50\,\mu$s: relative energy errors at the final parameters (means with the sample standard deviation in the last digits). Noiseless VQE: 20 runs of 20,000 iterations. Noisy VQE: noiseless relative error $\varepsilon_0 = (E_0 - E_g)/|E_g|$, noisy error, and normalized noise-induced shift $d = (E_{\lambda=1} - E_0)/|E_g|$; ZNE-assisted VQE: signed extrapolated error $(E_{\text{ZNE}} - E_g)/|E_g|$, noiseless error ZNE$^\dagger$, and $b = (E_{\text{ZNE}} - E_0)/|E_g|$; five runs of 10,000 iterations each. Mixed: mixed-canonical.

| | | Noiseless | Noisy VQE | | | ZNE-assisted VQE | | |
|---|---|---|---|---|---|---|---|---|
| $(L, K, \chi)$ | Ansatz | VQE | $\varepsilon_0$ | Noisy | $d$ | Extrap. | ZNE$^\dagger$ | $b$ |
| (3, 2, 8) | Ladder | 0.07346(14) | 0.08196(5) | 0.20586(7) | 0.12390(5) | 0.066(3) | 0.0716(13) | −0.006(3) |
| | Mixed | 0.0694(8) | 0.07531(3) | 0.19820(7) | 0.12288(9) | 0.0731(8) | 0.0703(16) | 0.003(2) |
| | Brick-wall | 0.021(13) | 0.032(3) | 0.141(4) | 0.108(4) | 0.031(24) | 0.028(25) | 0.003(2) |
| (4, 3, 8) | Ladder | 0.0010(2) | 0.0050(3) | 0.1802(19) | 0.175(2) | −0.002(4) | 0.0027(6) | −0.005(4) |
| | Mixed | 0.0022(2) | 0.0081(6) | 0.1819(5) | 0.1737(7) | 0.0144(13) | 0.0034(3) | 0.0110(13) |
| | Brick-wall | 0.0011(2) | 0.051(33) | 0.159(25) | 0.108(8) | 0.0086(16) | 0.0018(2) | 0.0068(15) |

**Table S7** Paired-protocol control for the single-processor layouts at $(L, K, \chi) = (3, 2, 8)$ and $(4, 3, 8)$, $N = 8$, $t_{2q} = 50\,\mu$s: for each layout, five stored initial-parameter sets (Sec. 3.B) shared by noiseless (20,000 iterations), noisy, and ZNE-assisted VQE (10,000 iterations); relative energy errors at the final parameters, means over the five runs with the sample standard deviation in the last digits. The mean encoding fidelities of the pretraining are 0.813–0.815 at $(3, 2, 8)$ and 0.989–0.992 at $(4, 3, 8)$. Noisy VQE: noiseless error $\varepsilon_0$, noisy error, and noise-induced shift $d = (E_{\lambda=1} - E_0)/|E_g|$; ZNE-assisted VQE: signed extrapolated error $(E_{\mathrm{ZNE}} - E_g)/|E_g|$, noiseless error ZNE$^\dagger$, and $b = (E_{\mathrm{ZNE}} - E_0)/|E_g|$. Mixed: mixed-canonical.

| $(L, K, \chi)$ | Ansatz | Noiseless VQE | Noisy VQE $\varepsilon_0$ | Noisy VQE Noisy | Noisy VQE $d$ | ZNE-assisted VQE Extrap. | ZNE-assisted VQE ZNE$^\dagger$ | ZNE-assisted VQE $b$ |
|---|---|---|---|---|---|---|---|---|
| $(3, 2, 8)$ | Ladder | 0.07329(2) | 0.0773(27) | 0.2094(21) | 0.132(5) | 0.0681(20) | 0.0721(13) | −0.0040(9) |
| | Mixed | 0.0689(7) | 0.07504(26) | 0.1987(7) | 0.1236(10) | 0.0732(5) | 0.06912(8) | 0.0041(5) |
| | Brick-wall | 0.0126(5) | 0.0313(24) | 0.1384(10) | 0.1072(14) | 0.043(28) | 0.041(30) | 0.0019(16) |
| $(4, 3, 8)$ | Ladder | 0.0007(3) | 0.0079(23) | 0.1807(30) | 0.173(3) | −0.003(4) | 0.0030(8) | −0.006(4) |
| | Mixed | 0.0020(6) | 0.0084(5) | 0.1813(17) | 0.1729(20) | 0.0142(10) | 0.0032(5) | 0.0110(7) |
| | Brick-wall | 0.0011(4) | 0.008(4) | 0.1506(25) | 0.1424(27) | 0.0076(24) | 0.0018(7) | 0.0058(18) |

**Table S8** Gate-error-dominated control at $(L, K, \chi) = (3, 2, 8)$, $N = 8$: noisy VQE with $p_d = 5 \times 10^{-3}$ per active qubit at $t_{2q} = 5\,\mu$s (GE) compared with the paired runs at $t_{2q} = 50\,\mu$s and $p_d = 10^{-4}$ of Table S7, from the same initial parameters (five runs; means with the sample standard deviation in the last digits). Noisy: noisy error at the final parameters; $\varepsilon_0$: noiseless error of the same parameters; $d$: noise-induced shift; $d_{\mathrm{GE}} - d_{50}$: paired difference of the shifts, positive in all five runs of every layout.

| | $50\,\mu$s, $p_d = 10^{-4}$ | | | $5\,\mu$s, $p_d = 5 \times 10^{-3}$ (GE) | | | |
|---|---|---|---|---|---|---|---|
| Ansatz | Noisy | $\varepsilon_0$ | $d$ | Noisy | $\varepsilon_0$ | $d$ | $d_{\mathrm{GE}} - d_{50}$ |
| Ladder | 0.2094(21) | 0.0773(27) | 0.132(5) | 0.250(3) | 0.0844(7) | 0.165(4) | +0.0333(7) |
| Mixed-canonical | 0.1987(7) | 0.07504(26) | 0.1236(10) | 0.2724(14) | 0.0830(5) | 0.1894(19) | +0.0658(23) |
| Brick-wall | 0.1384(10) | 0.0313(24) | 0.1072(14) | 0.221(10) | 0.053(16) | 0.168(8) | +0.061(9) |

**Table S9** Final relative energy errors after 10,000 iterations for the distributed architectures at $(L, K, \chi) = (3, 2, 8)$ for $(r_t, r_p) = (1, 1)$ and $(2, 2)$, with inter-QPU gate durations and biased-Pauli error probabilities equal to the intra-QPU values or twice those values, respectively. Values are arithmetic means over five runs, each initialized by an independent MPS pretraining run (with the initial parameters reused at both cost settings), with the sample standard deviation in the last digits given in parentheses.

| | Noisy | | ZNE | |
|---|---|---|---|---|
| $(r_t, r_p)$ | NN | All-to-all | NN | All-to-all |
| $(1, 1)$ | 0.20(3) | 0.183(12) | 0.026(7) | 0.021(4) |
| $(2, 2)$ | 0.245(23) | 0.242(6) | 0.048(13) | 0.066(19) |

## 3.G Sensitivity to the inter-QPU communication cost

To test how the uniform-cost assumption of the distributed simulations affects the Fig. 7 conclusions, we repeated the distributed noisy and ZNE-assisted calculations with inter-QPU two-qubit operations assigned twice the duration and error probability of intra-QPU ones. This section examines the twofold penalty $(2, 2)$ as a function of the layer number; the full cost scan of main-text Fig. 8, the selection of the two-QPU link, and the controls are described in Sec. 3.H. The penalty is parameterized by $r_t = t_{2\text{q,inter}}/t_{2\text{q,intra}}$ and $r_p = p_{d,\text{inter}}/p_{d,\text{intra}}$; the main-text baseline corresponds to $(r_t, r_p) = (1, 1)$, and the tested setting is $(2, 2)$, i.e. $t_{2\text{q,inter}} = 100\,\mu\text{s}$ and $p_{d,\text{inter}} = 2 \times 10^{-4}$ with every other Hamiltonian, ansatz, pretraining, optimizer, and noise setting unchanged. The inter-QPU duration enters both the active two-qubit thermal channel and the idle thermal channel of every scheduled sublayer containing an inter-QPU gate. Single-processor brick-wall circuits contain no inter-QPU operations and are unaffected. Each $(2, 2)$ ensemble uses five independent runs of 10,000 Adam iterations with fixed random-number seeds, each starting from the same pretrained initial parameters as the corresponding $(1, 1)$ run (Sec. 3.B); statistics of the final errors follow the main-text convention (after 10,000 iterations, mean $\pm$ sample SD).

Figure S4 and Tables S9–S10 summarize the results. The unmitigated layer sweep at $(r_t, r_p) = (2, 2)$ [Fig. S4(a)] preserves the corresponding $(1, 1)$ NN–all-to-all ordering at each total layer number. From $L = 3$ to $L = 4$ the all-to-all final error increases at both cost settings, from 0.183(12) to 0.194(9) at $(1, 1)$ and from 0.242(6) to 0.27(3) at $(2, 2)$, whereas the NN final error decreases slightly, from 0.20(3) to 0.186(2) and from 0.245(23) to 0.244(2). At $(L, K, \chi) = (3, 2, 8)$ [Fig. S4(b)], the penalty raises the unmitigated final error from 0.20(3) to 0.245(23) for NN and from 0.183(12) to 0.242(6) for all-to-all, and the final ZNE errors from 0.026(7) to 0.048(13) and from 0.021(4) to 0.066(19); all extrapolated trajectories remain above $E_g$ (Sec. 4).

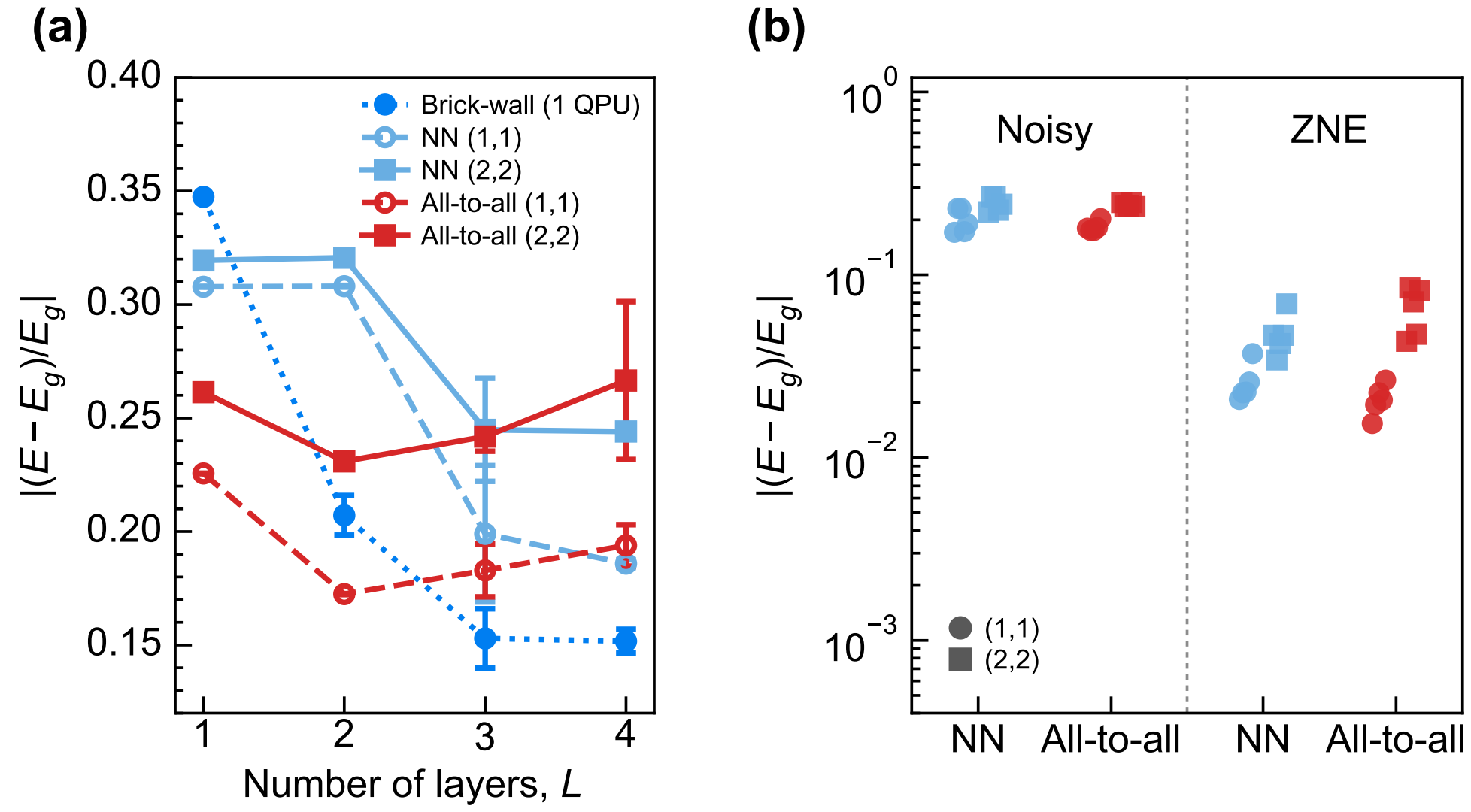


**Fig. S4** Sensitivity of the distributed comparison to the inter-QPU communication cost. (a) Unmitigated final relative energy error after 10,000 iterations versus the total layer number $L$ for $(r_t, r_p) = (1, 1)$ (open circles, dashed) and $(2, 2)$ (filled squares, solid); the single-QPU brick-wall series (dotted) contains no inter-QPU gates and is shown once. Markers and error bars denote the mean and sample standard deviation over five runs, each initialized by an independent MPS pretraining run; at $L = 1$ all pretraining runs reach the same encoding fidelity (to within $10^{-6}$) and the runs nearly coincide. (b) Comparison of final relative energy errors at $(L, K, \chi) = (3, 2, 8)$ for unmitigated and ZNE-assisted optimization; markers show the five individual runs; means and sample standard deviations are listed in Table S9.

**Table S10** Unmitigated final relative energy errors after 10,000 iterations versus the total layer number $L$ at $(r_t, r_p) = (2, 2)$ (means over five runs, each initialized by an independent MPS pretraining run; parentheses give the sample standard deviation in the last digits).

| | $L = 1$ | $L = 2$ | $L = 3$ | $L = 4$ |
|---|---|---|---|---|
| NN (2, 2) | 0.319(1) | 0.3206(4) | 0.245(23) | 0.244(2) |
| All-to-all (2, 2) | 0.2614(1) | 0.2309(6) | 0.242(6) | 0.27(3) |

## 3.H Inter-QPU cost scan, link selection, and reference calculations (Fig. 8)

Main-text Fig. 8 scans the common inter-QPU cost factor $r = r_t = r_p \in \{1, 2, 5, 10\}$ at the baseline two-qubit-gate duration $t_{2q} = 50\,\mu$s and at $t_{2q} = 5\,\mu$s for the three-QPU NN and all-to-all circuits, the two-QPU link circuit, and the single-QPU brick-wall, all at $(L, K, \chi) = (3, 2, 8)$ with the remaining noise settings of Table S4. As in the $(2, 2)$ study, $r_t$ scales the duration of every scheduled sublayer that contains an inter-QPU gate, and thereby the idle thermal-relaxation exposure of the qubits not acted on in that sublayer, whereas $r_p$ scales the biased-Pauli error probability of the inter-QPU gate. The simulations use five runs of 10,000 Adam iterations per ensemble with fixed random-number seeds and the main-text convention for reporting final errors. Each run is initialized by an independent MPS pretraining run; the resulting initial parameters are reused by the noiseless, noisy, and ZNE-assisted runs of the same circuit at all cost settings (Sec. 3.B), and for the brick-wall and three-QPU circuits the $r = 1$ ensembles at $50\,\mu$s are those of Fig. 7(c)–(f). The ZNE$^\dagger$ rows of Table S12 list the noiseless relative energy error at the final ZNE parameters, computed from the energy of the ideal circuit state prepared with those parameters (not of the noisy output state). At $t_{2q} = 5\,\mu$s and $r \leq 2$ the ensemble-mean difference between the extrapolated and the noiseless energy at the final parameters stays below $10^{-3}$ in relative error, whereas at $r \geq 5$ the extrapolated values exceed the noiseless ones by up to $5 \times 10^{-3}$ on average; at $50\,\mu$s the mean difference for the four circuits of Fig. 8 is 0.005–0.009 at $r = 1$ and reaches 0.06 at higher cost (Sec. 4). At $50\,\mu$s, $r = 1$, the mean ZNE$^\dagger$ errors of the three distributed circuits lie below that of the brick-wall, 0.012(3)–0.017(8) versus 0.0205(19); the two-QPU mean, 0.0200(18) at $r = 5$ and 0.036(3) at $r = 10$, stays below the brick-wall value up to $r = 5$ and exceeds it at $r = 10$. At $5\,\mu$s all three distributed circuits have lower mean ZNE$^\dagger$ errors than the brick-wall value of 0.0190(14) at every $r$, the NN circuit with a run-to-run spread of up to 0.010. The lower extrapolated errors in Fig. 8 are therefore accompanied by lower noiseless energy errors at the final ZNE parameters, although the extrapolated and noiseless errors differ at high communication cost.

*Link selection.* The two-QPU circuit cuts the chain into $\{0, \ldots, 3\}$ and $\{4, \ldots, 8\}$ and adds one inter-QPU block per layer between a communication qubit $a \in \{0, \ldots, 3\}$ and $b \in \{4, \ldots, 8\}$, so that its layer contains eight blocks, the same number as the brick-wall, at depth $D_2 = 3$ [Eq. (S12) with $M = 2$]. The link was selected in an initial noiseless VQE scan over the interaction pairs of the twisted Hamiltonian that connect the two parts and retained after the independent five-run scan reported here, which also includes the chain bond $(3, 4)$ and the non-interacting pair $(1, 6)$. Table S11 lists the final noiseless errors of the candidates over five pretraining runs each. The link $(3, 6)$ has the lowest mean final noiseless error among the two-QPU candidates, 0.0079(11), closely followed by $(1, 8)$, 0.0082(31); the brick-wall reaches 0.0173(42), the three-QPU NN circuit 0.015(10), and the all-to-all circuit 0.0050(18). The chain bond $(3, 4)$ gives 0.0193(15), and the interaction pairs $(0, 8)$ and $(3, 5)$ and the non-interacting pair $(1, 6)$ give higher means than the brick-wall: neither the summed graph distance of the interaction pairs nor the encoding fidelity predicts the final noiseless error, which is why the link was chosen by the scan.

**Table S11** Final noiseless VQE relative energy errors after 10,000 iterations for the two-QPU link candidates (partition $\{0,\ldots,3\} \mid \{4,\ldots,8\}$, one inter-QPU block per layer, $D_2 = 3$) and the reference circuits, at $(L, K, \chi) = (3, 2, 8)$, with one independent MPS pretraining run per VQE run and five runs per candidate (seeds distinct from those of Figs. 7 and 8; means with the sample standard deviation in the last digits in parentheses). "Pair" states whether the link is an interaction pair of the twisted Hamiltonian; $d_\Sigma$ is the summed graph distance of the eight interaction pairs on the coupling graph of the circuit; $F$ is the mean encoding fidelity of the five pretraining runs.

| Circuit | Link | Pair | $d_\Sigma$ | $F$ | Final noiseless error |
|---|---|---|---|---|---|
| 2 QPUs | $(3,6)$ | yes | 24 | 0.958 | 0.0079(11) |
| 2 QPUs | $(1,8)$ | yes | 24 | 0.805 | 0.0082(31) |
| 2 QPUs | $(2,7)$ | yes | 20 | 0.912 | 0.00996(22) |
| 2 QPUs | $(2,6)$ | yes | 21 | 0.951 | 0.0135(94) |
| 2 QPUs | $(1,7)$ | yes | 21 | 0.838 | 0.0169(50) |
| 2 QPUs | $(3,4)$ (chain bond, control) | no | 36 | 0.776 | 0.0193(15) |
| 2 QPUs | $(0,8)$ | yes | 29 | 0.803 | 0.0199(50) |
| 2 QPUs | $(3,5)$ | yes | 29 | 0.781 | 0.0249(60) |
| 2 QPUs | $(1,6)$ | no | 22 | 0.833 | 0.0304(43) |
| 1 QPU, brick-wall | – | – | 36 | 0.776 | 0.0173(42) |
| 3 QPUs, NN | $(0,3),(3,6)$ | – | 26 | 0.860 | 0.015(10) |
| 3 QPUs, all-to-all | $(0,3),(3,6),(0,6)$ | – | 21 | 0.965 | 0.0050(18) |

*Nearest-neighbor Heisenberg reference.* To measure the cost of distribution when no connectivity can be gained, the same circuits were run for the nearest-neighbor Heisenberg chain on nine qubits, whose interaction pairs $(i, i+1)$ are all nearest neighbors in the qubit order; the twisted model is a site permutation of this chain and shares its spectrum. The target MPS is a $\chi = 8$ DMRG state (relative energy error $7 \times 10^{-6}$, weight 0.99999 in the ground doublet). The circuits are the single-QPU brick-wall, the two-QPU circuit with the chain bond $(3,4)$ as its link, and three-QPU circuits with partition $\{0,1,2\} \mid \{3,4,5\} \mid \{6,7,8\}$ and communication qubits 2, 3, and 6, so that the NN links $(2,3)$ and $(3,6)$ restore one of the two cut bonds while the other, $(5,6)$, cannot be restored with one communication qubit per QPU; the all-to-all circuit adds $(2,6)$. Table S13 lists the results. Since no long-range interaction can be gained, every distributed layout has a higher unmitigated energy error than the brick-wall, with errors that increase with the two-qubit-block depth ($D_2 = 2, 3, 4, 5$ for the brick-wall, link, NN, and all-to-all circuits), at both gate durations, and the gap widens with $r$; the noiseless energies at the final ZNE parameters follow the same order at $50\,\mu$s, whereas at $5\,\mu$s the NN and all-to-all values differ by less than $10^{-3}$. The separately optimized noiseless VQE errors lie below $3 \times 10^{-3}$. In the ZNE-assisted runs, the extrapolated estimator crosses $E_g$ in 31 of the 55 runs (Sec. 4); the table therefore lists the noiseless relative energy error at the final ZNE parameters together with the signed extrapolated error, whose absolute value does not measure the state quality here.

**Table S12** Final relative energy errors after 10,000 iterations for the inter-QPU cost scan of main-text Fig. 8 (one independently pretrained initial parameter set per run, reused across all methods and cost settings of a circuit; means over five runs with the sample standard deviation in the last digits in parentheses). BW: single-QPU brick-wall ($D_2 = 2$, no inter-QPU gates, listed once per duration); NN and All: three-QPU NN ($D_2 = 4$) and all-to-all ($D_2 = 5$) circuits; Link: two-QPU circuit with the link $(3, 6)$ ($D_2 = 3$); Control: the same two-QPU circuit with the chain bond $(3, 4)$ (run at $r = 1$ only). ZNE$^\dagger$: noiseless relative energy error of the circuit at the final ZNE parameters. The last row gives the final noiseless VQE error of each circuit from the same initial parameters.

| $t_{2q}$ | Method | $r$ | BW | NN | All | Link | Control |
|---|---|---|---|---|---|---|---|
| 50 $\mu$s | Noisy | 1 | 0.153(13) | 0.20(3) | 0.183(12) | 0.133(8) | 0.1841(8) |
| | | 2 | | 0.245(23) | 0.242(6) | 0.1709(7) | |
| | | 5 | | 0.343(4) | 0.346(14) | 0.243(10) | |
| | | 10 | | 0.4003(16) | 0.443(7) | 0.326(13) | |
| | ZNE | 1 | 0.0264(21) | 0.026(7) | 0.021(4) | 0.0205(18) | 0.0345(10) |
| | | 2 | | 0.048(13) | 0.066(19) | 0.0271(4) | |
| | | 5 | | 0.11(3) | 0.155(30) | 0.060(3) | |
| | | 10 | | 0.211(20) | 0.204(28) | 0.096(29) | |
| | ZNE$^\dagger$ | 1 | 0.0205(19) | 0.017(8) | 0.012(3) | 0.0155(11) | 0.0211(10) |
| | | 2 | | 0.032(13) | 0.029(7) | 0.0154(7) | |
| | | 5 | | 0.0534(14) | 0.13(6) | 0.0200(18) | |
| | | 10 | | 0.192(6) | 0.18(3) | 0.036(3) | |
| 5 $\mu$s | Noisy | 1 | 0.0412(17) | 0.043(9) | 0.0347(16) | 0.0337(9) | 0.0466(10) |
| | | 2 | | 0.051(6) | 0.0489(11) | 0.0392(9) | |
| | | 5 | | 0.081(7) | 0.0875(21) | 0.05294(16) | |
| | | 10 | | 0.16(4) | 0.139(6) | 0.0768(19) | |
| | ZNE | 1 | 0.0185(13) | 0.012(9) | 0.0039(20) | 0.0077(14) | 0.0158(22) |
| | | 2 | | 0.012(9) | 0.0046(16) | 0.0078(14) | |
| | | 5 | | 0.017(10) | 0.0071(25) | 0.0081(14) | |
| | | 10 | | 0.020(9) | 0.013(3) | 0.0096(16) | |
| | ZNE$^\dagger$ | 1 | 0.0190(14) | 0.012(9) | 0.0045(21) | 0.0081(15) | 0.0160(23) |
| | | 2 | | 0.012(9) | 0.0047(17) | 0.0081(15) | |
| | | 5 | | 0.016(10) | 0.0056(19) | 0.0081(15) | |
| | | 10 | | 0.014(8) | 0.008(4) | 0.0084(20) | |
| – | Noiseless | – | 0.020(6) | 0.017(10) | 0.0036(5) | 0.0081(15) | 0.018(6) |

*Initialization sensitivity.* Repeating the 50 $\mu$s, $r = 1$ simulations from random initial parameters (no MPS pretraining; five runs) gives final unmitigated errors of 0.209(12) for the brick-wall, 0.1570(3) for NN, 0.160(21) for all-to-all, and 0.131(8) for the link circuit, compared with 0.153(13), 0.20(3), 0.183(12), and 0.133(8) from the pretrained runs (the two sets do not share initial states), and final noiseless errors

**Table S13** Nearest-neighbor Heisenberg reference: final relative energy errors after 10,000 iterations for the nine-qubit Heisenberg chain whose couplings are all nearest-neighbor in the qubit order (one independently pretrained initial parameter set per run, reused across all methods and cost settings of a circuit; five runs per entry; means with the sample standard deviation in the last digits in parentheses). BW: single-QPU brick-wall ($D_2 = 2$); $(3, 4)$: two-QPU circuit linked by the chain bond ($D_2 = 3$); NN and All: three-QPU circuits with communication qubits 2, 3, 6 ($D_2 = 4$ and 5). ZNE$^\dagger$: noiseless relative energy error at the final ZNE parameters; the fraction after each value is the number of the five runs whose final extrapolated energy lies below $E_g$. ZNE, signed: mean of $(E_{\mathrm{ZNE}} - E_g)/|E_g|$ at the final iteration. The signed ZNE rows report means only.

| $t_{2q}$ | Method | $r$ | BW | $(3, 4)$ | NN | All |
|---|---|---|---|---|---|---|
| – | Noiseless | – | 0.00016(6) | 0.00017(5) | 0.00258(25) | 0.00170(26) |
| 50 $\mu$s | Noisy | 1 | 0.0502(3) | 0.0717(10) | 0.0989(4) | 0.1206(4) |
| | | 2 | | 0.0924(8) | 0.1388(7) | 0.1768(8) |
| | ZNE$^\dagger$ | 1 | 0.0020(6) (5/5) | 0.0029(6) (5/5) | 0.0072(6) (1/5) | 0.016(4) (4/5) |
| | | 2 | | 0.0052(14) (1/5) | 0.016(3) (4/5) | 0.0193(7) (5/5) |
| | ZNE, signed | 1 | −0.0037 | −0.0017 | +0.0012 | −0.0138 |
| | | 2 | | +0.0003 | −0.0115 | −0.0264 |
| 5 $\mu$s | Noisy | 1 | 0.0108(4) | 0.0152(5) | 0.0211(5) | 0.0243(5) |
| | ZNE$^\dagger$ | 1 | 0.00023(6) (4/5) | 0.00022(5) (2/5) | 0.00257(28) (0/5) | 0.00180(16) (0/5) |
| | ZNE, signed | 1 | −0.00007 | +0.00003 | +0.0022 | +0.0014 |

of 0.016(6), 0.016(13), 0.008(6), and 0.011(5). For each circuit, every pretrained run reaches a relative error of 0.30 earlier than every randomly initialized run (within 65 iterations, compared with 62–268). For the distributed circuits, pretraining therefore accelerates the initial decrease in energy error but does not lower the final unmitigated error, whereas for the brick-wall it does; the relative ordering with the brick-wall circuit changes under random initialization. With random initialization the all-to-all ZNE estimator crosses $E_g$ in four of five runs.

## 3.I Pretrained start and trained circuits (Table 2)

Main-text Table 2 compares, for each circuit, the pretrained start with the circuits obtained from it by noiseless, noisy, and ZNE-assisted VQE. The pretrained initial circuit at $\boldsymbol{\theta}_{\mathrm{enc}}$ consists of the MPS-pretrained layers and the appended near-identity layer, and its energies are evaluated before VQE optimization; the noiseless, noisy, and ZNE-assisted runs of a circuit start from the same five parameter sets at every noise setting (Sec. 3.B). Noiseless, noisy, and ZNE-assisted VQE are evaluated with the noiseless, noisy, and extrapolated energies, respectively, and the noiseless energies at the noisy- and ZNE-trained parameters serve as diagnostics. Table S14 extends Table 2 to all settings of Fig. 8.

We write the noisy relative energy error as $(E_{\lambda=1} - E_g)/|E_g| = (E_0 - E_g)/|E_g| + d$, with the normalized noise-induced shift $d = (E_{\lambda=1} - E_0)/|E_g|$, and compare five-run means at the pretrained start and after noisy training; percentage reductions are relative to the means at the pretrained start, and contributions to the decrease of the noisy error are ratios of the corresponding mean decreases. This decomposition of the

change from the pretrained initial circuit to the trained circuit does not separate the two contributions during the optimization, since both depend on the same parameters. Noisy training lowers the noisy error relative to the pretrained start in all 140 runs (including the chain-bond control) and lowers $d$ in 129 of the 130 runs of Fig. 8. At $50\,\mu$s and $r = 1$, $d$ decreases by 17% for the brick-wall and by 34%, 46%, and 44% for the link, NN, and all-to-all circuits, and the reduction of $d$ accounts for 15% of the decrease of the noisy error for the brick-wall and for 53%, 56%, and 76% for the link, NN, and all-to-all circuits, respectively; for NN the runs differ widely. At $5\,\mu$s and $r = 1$, the decrease of the noiseless error accounts instead for 89–96% of the decrease of the noisy error in the distributed circuits, and the mean shift of the brick-wall remains essentially unchanged. At higher communication cost the distributed circuits reduce $d$ by up to 84% ($50\,\mu$s, $r = 10$), while their noiseless error at the trained parameters can exceed that of the pretrained start. For the nearest-neighbor Heisenberg reference, where the links provide no additional coverage of Hamiltonian bonds relative to the brick-wall, the pretrained initial circuits have mean noiseless errors of 0.011–0.020, the mean noise-induced shift decreases by 61–68% at $50\,\mu$s in all circuits, and the brick-wall has the lowest mean noisy error after noisy training at every tested setting.

A lower error than the brick-wall at the pretrained start does not ensure a lower error after training. At $50\,\mu$s, the link circuit has a lower noisy error than the brick-wall at the pretrained start for $r \leq 2$ but after noisy training only at $r = 1$, and a lower mean extrapolated error at the pretrained start up to $r = 5$ but after ZNE-assisted training only at $r = 1$. At $5\,\mu$s, the link and all-to-all circuits have lower noisy errors at the pretrained start up to $r = 10$ and $r = 5$, respectively, but after noisy training only up to $r = 2$ and $r = 1$; after ZNE-assisted training, their mean extrapolated errors remain below the brick-wall value at every tested $r$.

## 3.J Coupling coverage without a fixed qubit order: a tree-coupled Hamiltonian

In the twisted model the qubit order is fixed by the MPS, and the inter-QPU links of the distributed circuits realize interaction pairs that are long-range in that order. To test whether additional coupling coverage improves the noisy performance when the qubit order of a single chain can be chosen freely, we use a nine-qubit Heisenberg model on a tree, a central site joined to three arms of three, three, and two sites (eight bonds). No chain of nearest-neighbor blocks covers all eight bonds; the best qubit orders cover seven. The three-QPU circuit with partition $\{0, 1, 2\} \,|\, \{3, 4, 5\} \,|\, \{6, 7, 8\}$ and links $(2, 3)$ and $(3, 6)$ in a suitable qubit order covers all eight bonds with the same number of blocks (24) and parameters (360) as the chain circuits. As chain references we use the seven-bond qubit order with the lowest final noiseless VQE error in a separate scan over the twelve inequivalent orders, evaluated with independent initial parameters, and the chain with the qubit order of the tree circuit. The targets are exact ground states (sector $S^z = +1/2$) with maximum bond dimension 16, the circuits use $(L, K) = (3, 2)$, and the pretraining follows Sec. 3.B.

Table S15 lists the results. At the pretrained start, the tree circuit is more accurate than both chains and at $5\,\mu$s and $r = 1$ also has a lower mean noisy error, but its noise-induced shift is larger. After noiseless VQE the tree circuit and the same-order

**Table S14** Pretrained start and trained circuits at all settings of Fig. 8 (means over five runs with the sample standard deviation in the last digits). The noiseless errors at the pretrained start do not depend on the noise setting: 0.169(17) (BW), 0.0756(4) (link), 0.16(3) (NN), 0.072(7) (all-to-all), and 0.1605(4) (chain bond). $d$: normalized noise-induced shift at the pretrained start and at the noisy-trained parameters. Noiseless (Noisy VQE): noiseless relative energy error at the noisy-trained parameters; at $50\,\mu$s and $r = 1$ these are the Noisy-VQE entries marked * in Table 2. The brick-wall and chain-bond rows refer to $r = 1$.

| $t_{2q}$ ($\mu$s) | $r$ | Circuit | Pretrained start Noisy | Pretrained start ZNE | Pretrained start $d$ | Noisy VQE Noisy | Noisy VQE Noiseless | Noisy VQE $d$ | ZNE VQE ZNE |
|---|---|---|---|---|---|---|---|---|---|
| 50 | 1 | BW | 0.303(12) | 0.178(17) | 0.134(5) | 0.153(13) | 0.041(26) | 0.112(13) | 0.0264(21) |
| 50 | 1 | Link | 0.2420(15) | 0.0907(5) | 0.1663(15) | 0.133(8) | 0.024(3) | 0.109(5) | 0.0205(18) |
| 50 | 1 | NN | 0.375(29) | 0.19(3) | 0.216(6) | 0.20(3) | 0.08(6) | 0.12(3) | 0.026(7) |
| 50 | 1 | All-to-all | 0.333(16) | 0.123(10) | 0.261(9) | 0.183(12) | 0.037(5) | 0.146(7) | 0.021(4) |
| 50 | 1 | Chain bond | 0.3529(26) | 0.1846(11) | 0.1924(25) | 0.1841(8) | 0.0340(7) | 0.1500(14) | 0.0345(10) |
| 50 | 2 | Link | 0.2899(22) | 0.1056(7) | 0.2142(22) | 0.1709(7) | 0.03018(19) | 0.1407(9) | 0.0271(4) |
| 50 | 2 | NN | 0.459(27) | 0.24(3) | 0.300(10) | 0.245(23) | 0.09(6) | 0.15(4) | 0.048(13) |
| 50 | 2 | All-to-all | 0.448(21) | 0.204(16) | 0.375(14) | 0.242(6) | 0.054(5) | 0.188(10) | 0.066(19) |
| 50 | 5 | Link | 0.413(4) | 0.1764(18) | 0.338(4) | 0.243(10) | 0.039(4) | 0.204(6) | 0.060(3) |
| 50 | 5 | NN | 0.640(21) | 0.428(29) | 0.481(18) | 0.343(4) | 0.17(4) | 0.18(4) | 0.11(3) |
| 50 | 5 | All-to-all | 0.671(26) | 0.47(3) | 0.599(19) | 0.346(14) | 0.13(5) | 0.22(4) | 0.155(30) |
| 50 | 10 | Link | 0.567(6) | 0.329(4) | 0.491(6) | 0.326(13) | 0.070(7) | 0.256(9) | 0.096(29) |
| 50 | 10 | NN | 0.803(15) | 0.664(22) | 0.644(25) | 0.4003(16) | 0.29792(6) | 0.1023(16) | 0.211(20) |
| 50 | 10 | All-to-all | 0.838(26) | 0.72(4) | 0.766(19) | 0.443(7) | 0.26(5) | 0.18(4) | 0.204(28) |
| 5 | 1 | BW | 0.189(16) | 0.169(17) | 0.0201(8) | 0.0412(17) | 0.0211(21) | 0.0201(4) | 0.0185(13) |
| 5 | 1 | Link | 0.0987(5) | 0.0758(4) | 0.02305(19) | 0.0337(9) | 0.0136(12) | 0.0200(7) | 0.0077(14) |
| 5 | 1 | NN | 0.19(3) | 0.16(3) | 0.0299(8) | 0.043(9) | 0.019(11) | 0.0236(20) | 0.012(9) |
| 5 | 1 | All-to-all | 0.108(8) | 0.072(7) | 0.0359(14) | 0.0347(16) | 0.0067(17) | 0.0280(3) | 0.0039(20) |
| 5 | 1 | Chain bond | 0.1879(5) | 0.1605(4) | 0.0274(3) | 0.0466(10) | 0.0214(13) | 0.0251(6) | 0.0158(22) |
| 5 | 2 | Link | 0.1053(5) | 0.0759(4) | 0.02964(28) | 0.0392(9) | 0.0136(13) | 0.0256(6) | 0.0078(14) |
| 5 | 2 | NN | 0.20(3) | 0.16(3) | 0.0429(12) | 0.051(6) | 0.017(8) | 0.0335(23) | 0.012(9) |
| 5 | 2 | All-to-all | 0.127(9) | 0.073(7) | 0.0551(23) | 0.0489(11) | 0.0078(7) | 0.0411(4) | 0.0046(16) |
| 5 | 5 | Link | 0.1247(7) | 0.0762(4) | 0.0491(5) | 0.05294(16) | 0.01528(15) | 0.03766(25) | 0.0081(14) |
| 5 | 5 | NN | 0.24(3) | 0.16(3) | 0.0805(25) | 0.081(7) | 0.020(6) | 0.0607(8) | 0.017(10) |
| 5 | 5 | All-to-all | 0.182(12) | 0.077(7) | 0.110(5) | 0.0875(21) | 0.014(6) | 0.073(5) | 0.0071(25) |
| 5 | 10 | Link | 0.1560(10) | 0.0777(4) | 0.0804(10) | 0.0768(19) | 0.0171(14) | 0.0597(30) | 0.0096(16) |
| 5 | 10 | NN | 0.30(3) | 0.17(3) | 0.139(5) | 0.16(4) | 0.08(7) | 0.084(23) | 0.020(9) |
| 5 | 10 | All-to-all | 0.264(15) | 0.096(9) | 0.192(8) | 0.139(6) | 0.026(5) | 0.113(11) | 0.013(3) |

chain reach 0.000611(7) and 0.00059(8), both below the selected chain's 0.00103(12). In the scan, four of the five runs of the same-order chain ended below every run of the selected chain, but one run ended at a relative error of $3 \times 10^{-3}$ and raised its mean above that of the selected chain; with the independent initial parameters of this comparison, the same-order chain reaches the lower mean error. After noisy VQE, the tree circuit has higher mean unmitigated errors than both chains at every tested setting (the chains contain no inter-QPU gates, so the tree circuit at $r = 2$ is compared with the chains at $r = 1$). At $50\,\mu$s, noisy training raises the noiseless error of the tree circuit from 0.0116(5) to 0.0211(2) at $r = 1$ and 0.0292(2) at $r = 2$, whereas it lowers those of the chains to 0.0091(4) (same order) and 0.0086(7) (selected). After ZNE-assisted optimization the noiseless error at the final parameters at $5\,\mu$s exceeds that of the same-order chain by only $3 \times 10^{-5}$ on average and lies below that of the selected chain, whereas at $50\,\mu$s both chains have lower means. Coverage of an

additional coupling therefore does not by itself lower the noisy energy at equal block number. The ZNE estimates of the tree circuit lie below $E_g$ in all 20 runs, and those of the chains at 50 $\mu$s in all 10 (Sec. 4).

**Table S15** Tree-coupled Hamiltonian: pretrained start and trained circuits (means over five runs with the sample standard deviation in the last digits). Tree: three-QPU circuit covering all eight bonds; Chain (same order): chain circuit with the qubit order of the tree circuit; Chain (selected): the chain selected by the noiseless scan. Start: pretrained start. ZNE VQE: signed relative error $(E_{\mathrm{ZNE}} - E_g)/|E_g|$ of the final extrapolated energy, with the number of runs below $E_g$; ZNE$^\dagger$: noiseless relative energy error at the final ZNE parameters (diagnostic).

| $t_{2q}$ | Stage | $r$ | Tree | Chain (same order) | Chain (selected) |
|---|---|---|---|---|---|
| – | Start, noiseless | – | 0.0116(5) | 0.0227(5) | 0.027(3) |
| – | Noiseless VQE | – | 0.000611(7) | 0.00059(8) | 0.00103(12) |
| 50 $\mu$s | Start, noisy | 1 | 0.2219(21) | 0.148(4) | 0.1500(23) |
| | | 2 | 0.307(4) | | |
| | Noisy VQE | 1 | 0.0810(4) | 0.0605(6) | 0.0612(8) |
| | | 2 | 0.1151(9) | | |
| | ZNE VQE | 1 | −0.0107(20) (5/5) | −0.0023(11) (5/5) | −0.0030(6) (5/5) |
| | | 2 | −0.0250(21) (5/5) | | |
| | ZNE$^\dagger$ | 1 | 0.0090(17) | 0.0036(15) | 0.0031(4) |
| | | 2 | 0.0177(17) | | |
| 5 $\mu$s | Start, noisy | 1 | 0.0400(5) | 0.0412(10) | 0.0455(29) |
| | | 2 | 0.0525(7) | | |
| | Noisy VQE | 1 | 0.0162(2) | 0.0130(7) | 0.0130(6) |
| | | 2 | 0.0234(4) | | |
| | ZNE VQE | 1 | −0.000101(26) (5/5) | +0.00027(7) (0/5) | +0.00065(7) (0/5) |
| | | 2 | −0.00012(6) (5/5) | | |
| | ZNE$^\dagger$ | 1 | 0.000724(4) | 0.00070(9) | 0.00102(14) |
| | | 2 | 0.000732(9) | | |

# 4 Zero-noise extrapolation

The ZNE results in Figs. 6–8 use global unitary folding [21] followed by quadratic Richardson extrapolation [22, 23]. If

$$U(\boldsymbol{\theta}) = G_m(\theta_m) \cdots G_2(\theta_2) G_1(\theta_1) \tag{S19}$$

is the ordered ideal gate sequence, the folded circuits are

$$U^{(1)} = U, \tag{S20}$$
$$U^{(3)} = U\, U^\dagger U, \tag{S21}$$
$$U^{(5)} = U\, U^\dagger U\, U^\dagger U. \tag{S22}$$

Equivalently, for odd fold factor $\lambda = 2n + 1$,

$$U^{(\lambda)} = U \left(U^\dagger U\right)^n. \tag{S23}$$

The inverse circuit applies the same decomposed gates in reverse order, with each single-qubit rotation and Ising interaction replaced by its inverse. In the absence of noise all folded circuits implement the same logical unitary. In the simulations, every additional forward or inverse gate occurrence is followed by the corresponding noisy channel and idle qubits receive thermal relaxation for the added sublayer duration, so $\lambda$ should be interpreted as a fold factor rather than an exact physical noise-strength multiplier.

Let $Q_\lambda^{(s)}(t)$ denote the folded-circuit energy or loss value for run $s$ at optimization iteration $t$ and fold $\lambda \in \{1, 3, 5\}$. The zero-noise value is estimated independently for each matched run and iteration by evaluating the quadratic interpolant at $\lambda = 0$:

$$Q_{\mathrm{ZNE}}^{(s)}(t) = \frac{15}{8} Q_1^{(s)}(t) - \frac{5}{4} Q_3^{(s)}(t) + \frac{3}{8} Q_5^{(s)}(t). \tag{S24}$$

The coefficients are the Lagrange interpolation weights for the points $\lambda = 1, 3, 5$; they sum to unity and reproduce any quadratic function of $\lambda$ exactly. In the ZNE-VQE runs, this Richardson-combined quantity is the optimized loss at every VQE iteration, not only a final post-processing step. Each matched ZNE objective evaluation uses the fold-1, fold-3, and fold-5 circuits, so it requires $1+3+5 = 9$ times as many gate occurrences as one unfolded circuit evaluation, before accounting for gradient-evaluation overhead. The simulations use deterministic density-matrix or expectation evaluation, so shot noise is absent from the simulated trajectories; in finite-shot hardware experiments the negative Richardson coefficient can amplify estimator variance [24]. More broadly, fundamental limits on quantum error mitigation connect achievable accuracy to sampling overhead [25]. The estimator is applied to the folded-circuit energy or loss before conversion to relative energy error. The trajectories are shown as computed, without smoothing or fitting. Because Eq. (S24) extrapolates outside the sampled fold interval and contains a negative coefficient, the ZNE curve is not guaranteed to lie within the folded value range, to be variational, or to improve every individual run.

This caveat is tested with the noiseless relative energy error at the final ZNE parameters (ZNE$^\dagger$ in Table S12). In the ensembles of Figs. 7 and 8 and of the chain-bond control (28 ensembles and 140 ZNE-assisted runs, of which Figs. 7 and 8 show 26 and 130) no extrapolated trajectory crosses $E_g$. The extrapolation error at the final parameters, $b = (E_{\mathrm{ZNE}} - E_0)/|E_g|$ with $E_0$ the noiseless energy of the same parameters, stays below $10^{-3}$ in magnitude in ensemble mean at $t_{2\mathrm{q}} = 5\,\mu$s and $r \leq 2$ (individual $|b|$ up to $1.3\times10^{-3}$; for the circuits of Fig. 8, $b < 0$ in 34 of the 35 runs) and reaches $5\times10^{-3}$ at $r \geq 5$ (individual runs $7.7\times10^{-3}$); at $50\,\mu$s, for the four circuits of Fig. 8, the ensemble means range from 0.005 to 0.009 at $r = 1$ and reach 0.06 at $r \geq 5$ (individual runs up to 0.12), so larger biases of the estimator occur at the higher-cost settings. For the nearest-neighbor Heisenberg reference, whose noiseless VQE errors lie below $3 \times 10^{-3}$, the same estimator crosses $E_g$ in 31 of 55 runs (Table S13), so that comparison uses unmitigated energies and the exact energy at the final ZNE parameters. With random initialization, the all-to-all ZNE estimator at $50\,\mu$s, $r = 1$, crosses $E_g$ in four of five runs (Sec. 3.H).

*Extrapolation error under ZNE-assisted training.* Because the extrapolated energy is the minimized quantity, the optimization can lower it partly by lowering the extrapolation error itself. For the circuits of Fig. 8 and the chain-bond control at $50\,\mu$s, $r = 1$, the mean normalized extrapolation error $b$ is $+0.008$ to $+0.051$ at the pretrained start and $+0.005$ to $+0.013$ after ZNE-assisted training. For the tree-coupled Hamiltonian (Sec. 3.J) at $50\,\mu$s, $b$ of the tree circuit changes from $+0.026$ at the pretrained start to $-0.020$ after ZNE-assisted training at $r = 1$ and from $+0.066$ to $-0.043$ at $r = 2$, and $b$ of both chains is $-0.006$ after ZNE-assisted training. In all five tree-circuit runs at $50\,\mu$s, $r = 1$ (four of five at $r = 2$), the extrapolated energy decreases from 1,000 to 10,000 iterations while the noiseless energy at the same parameters increases, and 30 of the 40 ZNE-assisted runs of the tree comparison end below $E_g$. For the single-processor ladder ansatz, the mean $b$ after ZNE-assisted training is $-0.005$ at $(L, K, \chi) = (4, 3, 8)$ (Fig. 6; negative in all five runs, three of which end below $E_g$) and $-0.006$ at $(3, 2, 8)$ (Table S6). Since the simulations contain no shot noise, the lowering of the extrapolated energy through the extrapolation error is systematic rather than statistical overfitting; ZNE$^\dagger$ gives the noiseless energy error of the optimized circuit. The source data (`data/table2` and `data/supplementary`) provide, for every run of the circuits of Fig. 8 and of the chain-bond, nearest-neighbor Heisenberg, and tree references, the energies at the pretrained start, the final noiseless, noisy, and extrapolated energies of the corresponding VQE runs, and the noiseless energies at the final parameters, and summarize by ensemble whether the ZNE-assisted runs fall below $E_g$ at the final iteration or at any point during optimization.

*Extrapolation at the noisy-trained parameters.* For the noisy-VQE runs of Fig. 8 and the chain-bond control at $r = 1$, the quadratic estimator of Eq. (S24) was evaluated once at the final parameters (Table S16). The estimate lies above the noiseless energy of the same parameters in all 50 runs, by 0.007–0.022 in ensemble mean at $50\,\mu$s and by less than $5\times10^{-4}$ at $5\,\mu$s, and, on average for every circuit, above the extrapolated energy reached by ZNE-assisted training from the same initial parameters; at $50\,\mu$s the paired difference is positive in all five runs of each circuit of Fig. 8, whereas at $5\,\mu$s it is negative in one of the five runs for each of the brick-wall and all-to-all circuits. The

extrapolation therefore estimates the energy of the state reached by noisy training, with a residual bias at the baseline duration, but does not move the parameters toward those found by ZNE-assisted training.

**Table S16** Extrapolation at the noisy-trained parameters for the circuits of Fig. 8 and the chain-bond control at $r = 1$ (five runs per entry; means with the sample standard deviation in the last digits). Noisy VQE: unmitigated error, the quadratic ZNE estimate of Eq. (S24) evaluated once at the final parameters (Extrapolated), and the noiseless error of the same parameters. ZNE-assisted VQE: final extrapolated error and ZNE$^\dagger$ from the same initial parameters (Table S12).

| | | Noisy VQE | | | ZNE-assisted VQE | |
|---|---|---|---|---|---|---|
| $t_{2q}$ | Circuit | Noisy | Extrapolated | Noiseless | Extrap. | ZNE$^\dagger$ |
| 50 $\mu$s | 1 QPU, brick-wall | 0.153(13) | 0.048(24) | 0.041(26) | 0.0264(21) | 0.0205(19) |
| | 2 QPUs, link $(3, 6)$ | 0.133(8) | 0.034(4) | 0.024(3) | 0.0205(18) | 0.0155(11) |
| | 3 QPUs, NN | 0.20(3) | 0.10(6) | 0.08(6) | 0.026(7) | 0.017(8) |
| | 3 QPUs, all-to-all | 0.183(12) | 0.059(9) | 0.037(5) | 0.021(4) | 0.012(3) |
| | 2 QPUs, chain bond $(3, 4)$ | 0.1841(8) | 0.0491(7) | 0.0340(7) | 0.0345(10) | 0.0211(10) |
| 5 $\mu$s | 1 QPU, brick-wall | 0.0412(17) | 0.0213(20) | 0.0211(21) | 0.0185(13) | 0.0190(14) |
| | 2 QPUs, link $(3, 6)$ | 0.0337(9) | 0.0139(12) | 0.0136(12) | 0.0077(14) | 0.0081(15) |
| | 3 QPUs, NN | 0.043(9) | 0.020(11) | 0.019(11) | 0.012(9) | 0.012(9) |
| | 3 QPUs, all-to-all | 0.0347(16) | 0.0069(18) | 0.0067(17) | 0.0039(20) | 0.0045(21) |
| | 2 QPUs, chain bond $(3, 4)$ | 0.0466(10) | 0.0217(13) | 0.0214(13) | 0.0158(22) | 0.0160(23) |

*Finite-shot cost.* At the final parameters of the noisy and ZNE-assisted runs of Fig. 8 at $r = 1$, the exact single-shot variance of the energy was computed for measurement in three settings (all qubits in $X$, $Y$, or $Z$) for the fold-1, fold-3, and fold-5 circuits. For a fixed total number of shots with optimal allocation, the variance of the ZNE estimate is 12.8–14.1 times that of the unmitigated estimate (16.1–17.3 times with equal allocation and 25–29 times for equal total circuit time). With the biases at these parameters, the mean squared error of the ZNE estimate relative to the noiseless energy at the same parameters falls below that of the unmitigated estimate above about 75–250 shots per energy estimate at $t_{2q} = 50\,\mu$s and about 1,500–6,000 shots at 5 $\mu$s (medians over the five runs of each circuit). These numbers describe the measurement cost at fixed parameters; optimization with shot noise was not simulated.


\end{document}